\documentclass[12pt]{conm-p-l}

\usepackage{amscd}
\usepackage{amssymb}
\usepackage{amsfonts}
\usepackage{graphicx}
\usepackage{amsmath}
\input{amssym.def}

\newtheorem{theorem}{Theorem}
\newtheorem{proposition}{Proposition}

\newtheorem{lemma}{Lemma}
\newtheorem{remark}{Remark}

\theoremstyle{remark}

 \numberwithin{equation}{section}
\newcommand{\re}{\text{\rm Re }}

\newcommand{\Sb}{\text{\bf S}}
\newcommand{\Diff}{\text{\rm Diff }}

\newcommand{\Vect}{\text{\rm Vect }}
\newcommand{\const}{\text{\rm const}}
\DeclareMathOperator{\p}{\partial}

\begin{document}

\title[Conformal field theory and L\"owner-Kufarev evolution]{Conformal field theory and L\"owner-Kufarev evolution}
\author{Irina Markina}
\address{University of Bergen, Johannes Brunsgate 12, N-5008 Bergen, Norway}
\email{Irina.Markina@uib.no}
\author{Alexander Vasil'ev}
\address{University of Bergen, Johannes Brunsgate 12, N-5008 Bergen, Norway}
\email{Alexander.Vasiliev@uib.no}
\thanks{This work was completed with the support of the grants of the Norwegian Research Council \#177355/V30,  of  the
European Science Foundation Research Networking Programme HCAA, and of the NordForsk Network `Analysis and Applications'}

\begin{abstract}

One of the  important aspects in recent trends in complex analysis has been the increasing degree of cross-fertilization 
between the latter and mathematical physics  with great benefits to both subjects. Contour dynamics in the complex plane turned to be a meeting point for  complex  analysts, specialists in stochastic processes, and mathematical physicists. This was stimulated, first of all, by recent progress in understanding  structures in the classical and stochastic L\"owner evolutions, and in the Laplacian growth. The Virasoro algebra provides  a basic algebraic object in conformal field theory (CFT) so it was not surprising that it turned to play an important role of a structural skeleton for contour dynamics.
The present paper is a survey of recent progress in  the study of the CFT viewpoint on contour dynamics, in particular, we show how the Witt and Virasoro algebras are related with the stochastic L\"owner and classical L\"owner-Kufarev equations.

\end{abstract}

\maketitle

\section{Introduction}
Conformal field theory (CFT) in two dimensions has deep intrinsic connection to representation of infinite dimensional algebras. The Virasoro algebra is a vertex algebra, which appeared in early 1970's physics papers (see, e.g. \cite{Virasoro}) on string theory. Earlier in 1968 it was introduced by Gelfand and Fuchs \cite{GelfandFuchs} as a unique (up to isomorphisms) central extension of the algebra of vector fields on the unit circle. Later in 1980's, it became clear that the Virasoro algebra turned out to be a universal symmetry algebra in two-dimensional CFT.
The infinitesimal conformal transformations in the classical setup lead to an infinite dimensional algebra, called the Witt algebra. Turning to quantum field theories, the conformal anomaly, or Weyl anomaly, leads to the appearance of a nontrivial central charge. So the Witt algebra is modified by central extension to the Virasoro algebra. Infinite-dimensional algebras have also been used recently in the theory of exactly solvable models. For example, the Virasoro algebra plays a central role in the study of integrable systems, such as those associated to the KdV and other soliton hierarchies. The Virasoro algebra is intrinsically related to the KdV canonical structure where the Virasoro brackets become just the Magri brackets for the Miura transformations of  elements of the phase space of the KdV hierarchy (see, e.g., \cite{Faddeev, Gervais}). The lattice Virasoro algebra appears in the study of the Toda field theory and Toda integrable systems \cite{FaddeevTakhtajan, Inoue}. 

On the other hand, contour dynamics is a classical subject in complex analysis. One of the typical dynamics started from classical Hele-Shaw experiments in 1897.  This leads to a sample free boundary problem, known also as the Laplacian growth in two dimensions (see \cite{ GustVas, Vas}). Another one is the L\"owner evolution, see \cite{Loewner, Pommerenke2}. Last decade, the progress in the study of Laplacian growth has resulted in its integrable structure in terms of Richardson's moments, namely these moments form a dispersionless Toda hierarchy \cite{Mineev}. Richardson's moments are conserved quantities of the Laplacian growth. Recently, it was proved \cite{MPV} that the Virasoro generators in their co-vector representation are conserved quantities of the L\"owner-Kufarev evolution.  Friedrich and Werner \cite{FriedrichWerner}, and independently Bauer and Bernard \cite{BB} found relations between SLE (stochastic or Schramm-L\"owner evolution) and the highest weight representation of the Virasoro algebra. All these results encouraged us to conclude that the Virasoro algebra is a common structural basis for these and possibly other types of contour dynamics and we decided to present our overview of the development in this direction.

The survey is designed for a comprehensive reading of sufficiently wide mathematical audience, in particular graduate students. Therefore, we decided to include several parts well-known for specialists in mathematical physics but less known for analysts. 

 We start to show how the Virasoro algebra appears in CFT and in integrable models. Then we proceed with the classical L\"owner-Kufarev equation where the central extension does not appear and we work only with the Witt algebra. The central extension appears for the stochastic version of the L\"owner equation. Finally, we briefly describe connections between SLE and CFT.

This survey is partially based on a lecture given by the second author at the II Winter School in Complex Analysis and Operator Theory, held in Seville (Spain), February 5-9, 2008. He is thankful to the organizers for their hospitality. The authors want to express their gratitude to H\'el\`ene Airault, Roland Friedrich, Paul Malliavin,  Yurii Neretin, and Dmitri Prokhorov for many fruitful discussions on the subject during last years.
  
\section{CFT and Virasoro algebra}

There is a vast amount of standard references on basics of CFT, and classical and quantum strings, see e.g., \cite{Green, Polchinski}, but since the topic is presented in a lecture form for mathematics students, we take a risk to look briefly through a simplest example of a closed bosonic string with periodic boundary conditions in order to introduce the Virasoro algebra.

\subsection{Classical bosonic string}
We start with some basic definitions. A $p$-brane is  spatial evolution of a $p$-dimensional object, which is reduced to a point particle if $p=0$, a string if $p=1$, a membrane if $p=2$, etc., in some $D$-dimensional ambient spacetime $\mathfrak M$. The result of such evolution is called the worldline ($p=0$), worldsheet ($p=1$), or worldvolume otherwise.
We suppose $\mathfrak M$ to be a $D$-dimensional vector space endowed with the Lorentzian metric $ds_{\mathfrak M}$ of signature $(1,D-1)$;  that is
\[
ds^2_{\mathfrak M}=\eta_{\mu\nu}dx^{\mu}dx^{\nu}, \quad \eta_{\mu\nu}=\left(\begin{array}{cccc}
 -1 & 0 & \dots & 0 \\ 
 0 & 1 & \dots & 0 \\ 
 \dots & \dots & \dots & \dots \\ 
0 & 0 & \dots & 1
\end{array}\right).
\]
We consider the toy model of a closed bosonic string. In order to see its dynamics we construct dynamical variables $x^{\mu}(\sigma^{\alpha})$ where $\mu=0,1,\dots, D-1$ and $\sigma^0=\tau$, $\sigma^1=\sigma$ are coordinates on the canonical cylinder 
$C=\{\tau\in (-\infty,\infty), \sigma\in [0,2\pi)\}$. The intrinsic metric $ds_C$ on $C$ is given by
\[
ds^2_{C}=\eta_{\alpha\beta}d\sigma^{\alpha}d\sigma^{\beta}, \quad \eta_{\alpha\beta}=\left(\begin{array}{cc}
 -1 & 0 \\ 
 0 & 1  \\ 
\end{array}\right).
\]
We reserve the notation $\eta_{\mu\nu}$ for the metric in $D$-dimensional space and $\eta_{\alpha\beta}$ on the cylinder $C$.
The dynamical variables describe a $C^{\infty}$-smooth embedding of $C$ in $\mathfrak M$, and this allows us to consider the worldsheet ($ws$) as an embedded manifold in $\mathfrak M$. 
The induced metric $ds_{ws}$ on the worldsheet is given by
\[
ds^2_{ws}= \eta_{\mu\nu}\frac{\partial x^{\mu}}{\partial \sigma^{\alpha}}\frac{\partial x^{\nu}}{\partial \sigma^{\beta}}\,d\sigma^{\alpha}d\sigma^{\beta}=g_{\alpha \beta}\,d\sigma^{\alpha}d\sigma^{\beta}.
\]
The dynamical variables $x^{\mu}(\tau,\sigma)$ satisfy the equations of motion derived by making use of the {\it least action principle}. In Lagrangian mechanics, a system with a configuration space $\mathfrak N$
is characterized by its {\it Lagrangian} $L$, a smooth real-valued functional on the direct product of the tangent bundle $T\mathfrak N$ and the real axis. Let $P\mathfrak N$ stand for the path space, that is the space of all paths having the fixed initial point (`ini') and the end point (`end'). This space has a structure of an infinite-dimensional Fr\'echet manifold. Then, the {\it action functional} $S$ is a real-valued integral functional defined by $S=\int_{\mbox{\small ini}}^{\mbox{\small fin}}L dt$, where $t$ is a real curve parameter. Turning to the spacetime configuration space $\mathfrak M$ and a worldsheet in $\mathfrak M$ we work with the action functional $S=\int_{ws}\Phi(x)dx$ where the Lagrangian $L$ is included in to this spatial integral. So defined action has some physical and geometrical relevance. In classical mechanics, action, e.g.,  is the difference between kinetic and potential energies. The minimizing curve for the length functional and for the action given by the Lagrangian, which is the kinetic energy, coincide.
There is no a direct analogue to energy in the relativistic mechanics. Therefore, Nambu  in 1970 \cite{Nambu} and
 Got{\^o} in 1971 \cite{Goto} proposed to choose 
the area of the worldsheet as the simplest action. This action, known as the Nambu-Goto action, admits the form
\[
S_{NG}=-T\int\limits_{ws}dx=-T\int\limits_{C}d\sigma^2\sqrt{|\det g_{\alpha\beta}|},
\]
where $C$ is the canonical cylinder and the quantity $T$, the string tension, has dimension mass per unit length. Observe that (-) in front of the integral is due to the Lorentzian metric. The string possesses geodesic motion, i.e., the dynamical variables satisfy the Euler-Lagrange equation
\begin{equation}\label{EL1}
\frac{1}{\sqrt{|\det g_{\alpha\beta}|}}\frac{\partial}{\partial \sigma^{\alpha}}
\left(\sqrt{|\det g_{\alpha\beta}|}\,\,g^{\alpha\beta}\frac{\partial x^{\mu}}{\partial \sigma^{\beta}}\right)=0,\quad \mu=0,\dots,D-1,
\end{equation}
where $g^{\alpha\beta}$ and $g_{\alpha\beta}$ are obtained by  rising and lowering of indices with respect to the Lorentzian metric. The metric $g_{\alpha\beta}$ is defined on the worldsheet embedded into $\mathfrak M$. It  depends on the variables $x^{\mu}$, and  the equation \eqref{EL1} for
$x^{\mu}$ is highly non-linear. Moreover, the square root within the integral results in difficulties in further quantization.

To overcome these problems Polyakov  proposed to introduce an analogue of Lagrange multipliers in 1981 \cite{Polyakov}. In fact, he considered a variable metric
$ds$ of index 1 on the worldsheet as on a 2-D manifold, such that $ds^2=h_{\alpha\beta}(\tau,\sigma)\,d\sigma^{\alpha}d\sigma^{\beta}$, and  the Dirichlet integral
\[
S_{P}=-\frac{T}{2}\int\limits_{C}d\sigma^2\sqrt{|\det h_{\alpha\beta}|}\,\,h^{\alpha\beta}\frac{\partial x^{\mu}}{\partial\sigma^{\alpha}} \frac{\partial x^{\nu}}{\partial\sigma^{\beta}}
\]
as an action. The Euler-Lagrange equation (regarding to the variation of the dynamical variables) for the Polyakov action $S_P$ is formally the same as for the Nambu-Goto action
$S_{NG}$
\begin{equation}
\frac{1}{\sqrt{|\det h_{\alpha\beta}|}}\frac{\partial}{\partial \sigma^{\alpha}}
\left(\sqrt{|\det h_{\alpha\beta}|}\,\,h^{\alpha\beta}\frac{\partial x^{\mu}}{\partial \sigma^{\beta}}\right)=0,\label{EL}
\end{equation}
but $h$ does not depend on $x$ any longer. So \eqref{EL} are  linear equations for $x^{\mu}(\tau,\sigma)$. Moreover,  the Polyakov action is  quantized easier due to its linearity.

There are three  degrees of freedom in $h$ because it is symmetric. They can be removed by using the equation of motion for $h$
\begin{equation}\label{EL3}
\frac{\delta S_P}{\delta h^{\alpha\beta}}=0,
\end{equation}
where the left-hand side is a functional derivative. The equations (\ref{EL}--\ref{EL3}) perform the necessary conditions for minimizing the action. The two-dimensional energy-momentum tensor is defined as
\[
T_{\alpha\beta}=\frac{-2}{T}\frac{1}{\sqrt{|\det h_{\alpha\beta}|}}\frac{\delta S_P}{\delta h^{\alpha\beta}},
\]
and the equation of motion implies $T_{\alpha\beta}=0$. Moreover, $S_P=S_{NG}$ under $T_{\alpha\beta}=0$, whereas in general, $S_P\geq S_{NG}$.

A guiding principle in physics is that symmetries in the geometry of spacetime define the standard conservation laws. There is no single theorem describing such a connection, however this principle is sometimes referred to as the `Noether theorem', although the Noether 1918 theorem \cite{Noether} itself contains only a very partial statement of it.

By symmetries for the Polyakov action\index{Polyakov action} $S_P$ we mean transformations keeping the action invariant. They are:
\begin{itemize}

\item { Global symmetries} 
 \begin{itemize}

\item  Poincar\'e invariance 
\[
x^{\mu}\to x^{\mu}+b^{\mu};
\] 
\[
x^{\mu}\to x^{\mu}+\omega^{\mu}_{\nu}x^{\nu};
\]
where $\omega^{\mu}_{\nu}=-\omega^{\nu}_{\mu}$ are infinitesimal  Lorenzian transformations.

\end{itemize}

\item { Local symmetries}

 \begin{itemize}

\item reparametrization invariance by diffeomorphisms in 2-D: $\sigma^{\alpha}\to \tilde{\sigma}^{\alpha}(\tau,\sigma)$ implies the invariance of the area element
\[
d\tilde{\sigma}^2\sqrt{|\det \tilde{h}|}=d\sigma^2\sqrt{|\det h|}.
\]

\item Weyl rescaling
\[
h_{\alpha\beta} d{\sigma}^{\alpha}d{\sigma}^{\beta}\to e^{\rho(\tau,\sigma)} h_{\alpha\beta} d{\sigma}^{\alpha}d{\sigma}^{\beta}.
\]

\end{itemize}

\end{itemize}
Weyl rescaling leaves $\sqrt{|\det h|}\,\,h_{\alpha\beta}$ invariant. 

The Poincar\'e and reparametrization invariance of $S_P$ allow us to choose a
gauge in which the three independent components of $h$ are expressed with
just one function, typically the conformal flat gauge $h_{\alpha\beta}= e^{\rho(\tau,\sigma)} \eta_{\alpha\beta}$. Substituting
this in the Polyakov action  we obtain
\[
S_{P}=-\frac{T}{2}\int\limits_{C}d\sigma^2\,\,\eta^{\alpha\beta}\eta_{\mu\nu}\frac{\partial x^{\mu}}{\partial\sigma^{\alpha}} \frac{\partial x^{\nu}}{\partial\sigma^{\beta}},
\]
so the gauge fixed action is a quadratic functional with respect to $x$. Varying it with respect to $x$ we arrive at a free wave equation of motion 
\begin{equation}\label{wave}
\ddot{x^{\mu}}-(x^\mu)''=0,
\end{equation} 
where $\dot{x}$ is the $\tau$- and $x'$ is the $\sigma$- derivative.

Weyl invariance implies that $S_P$ does not depend on $\rho(\tau,\sigma)$. Therefore, the variation $\delta S_P/\delta \rho=h^{\alpha\beta}T_{\alpha\beta}=T_{\alpha}^{\alpha}$ vanishes that makes $T_{\alpha\beta}$  traceless. 

Keeping in mind that $T_{\alpha\beta} = 0$ one can derive the constraints
$T_{01}=T_{10}=\dot{x}_{\mu}(x_{\mu})'=0$ and 
\begin{equation}\label{constraint}
T_{00}=T_{11}=\frac{1}{2}(\dot{x}_{\mu}\dot{x}^{\mu}+x_{\mu}'(x^{\mu})')=0.
\end{equation}
This yields the equations
$(\dot{x}\pm x')^2=0$, which are called the Virasoro constraints.
The equation \eqref{wave} with the constraints \eqref{constraint}, subject to some boundary
conditions describes the motion of a bosonic string. 

Let us use periodic boundary conditions $x^{\mu}(\tau,\sigma)=x^{\mu}(\tau,\sigma+2\pi)$. The general solution
to the wave equation of motion is
\[
x^{\mu}(\tau,\sigma)=x^{\mu}_R(\tau-\sigma)+x^{\mu}_L(\tau+\sigma).
\]
Let us show that the center of mass of the worldsheet moves as a free particle.
Satisfying the boundary condition and
separating the linear part,
  we use the Fourier series representation of the solution as \[
x^{\mu}_R(\tau-\sigma)=\frac{1}{2}x_0^{\mu}+\frac{1}{2\pi T}(\tau-\sigma)p^{\mu}+\frac{i}{\sqrt{2\pi T}}\sum\limits_{n\neq 0}\frac{1}{n}\alpha_n^{\mu}e^{-in(\tau-\sigma)},
\]
\[
x^{\mu}_L(\tau+\sigma)=\frac{1}{2}x_0^{\mu}+\frac{1}{2\pi T}(\tau+\sigma)p^{\mu}+\frac{i}{\sqrt{2\pi T}}\sum\limits_{n\neq 0}\frac{1}{n}\beta_n^{\mu}e^{-in(\tau+\sigma)},
\]
where  $x^{\mu}_0$ is the center of mass and $p^{\mu}$ is the momentum.
The functions $x^{\mu}_R$ and $x^{\mu}_L$ are real, and hence, $\bar{\alpha}^{\mu}_n=\alpha^{\mu}_{-n}$ and $\bar{\beta}^{\mu}_n=\beta^{\mu}_{-n}$. These coefficients are known under the name of oscillators in physics.

The position of the center of mass of the string is calculated as
\[
X^{\mu}=\frac{1}{2\pi}\int_0^{2\pi}x^{\mu}(\tau,\sigma)d\sigma=x_0^{\mu}+\frac{p^{\mu}}{\pi T}\tau,
\]
that shows that the center of mass moves as a free particle starting from $x_0^{\mu}$.
The momentum of the center of mass is written
\[
P^{\mu}=\int_0^{2\pi}\Pi^{\mu}d\sigma=\frac{T}{2}\int_0^{2\pi}\dot{x}^{\mu}d\sigma=p^{\mu}.
\]

In classical Hamiltonian mechanics the action $S$ is a time-integral of the Lagrangian $S=\int_{t_0}^{t_1}L(x,\dot{x})dt$.  In our case the Lagrangian becomes
\[
L=-\frac{T}{2}\int_0^{2\pi}\eta^{\alpha\beta}\eta_{\mu\nu}\frac{\partial x^{\mu}}{\partial\sigma^{\alpha}} \frac{\partial x^{\nu}}{\partial\sigma^{\beta}}d\sigma.
\]
The Hamiltonian function is given by
\[
H=\int_{0}^{2\pi}(\dot{x}^{\mu}\Pi_{\mu})d\sigma-L=\frac{T}{2}\int_{0}^{2\pi}(\dot{x}^{\mu}\dot{x}_{\mu}+(x^{\mu})'x_{\mu}')d\sigma.
\]
Using an identity for the Kronecker delta
\[
\frac{1}{2\pi}\int_0^{2\pi}e^{i(n-m)\sigma}d\sigma=\delta_{n,m},
\]
we obtain the Hamiltonian function in terms of oscillators as
\[
H=\frac{1}{2}\sum\limits_{n\in\mathbb Z}(\alpha_{-n}\alpha_n+\beta_{-n}\beta_n),
\]
where we set $\alpha^{\mu}_0=\beta_0^{\mu}=\frac{1}{2\pi T}p^{\mu}$. 

The standard Heisenberg-Poisson-Dirac bracket  on phase coordinates looks like
\[
\{x^{\mu}(\tau,\sigma), \dot{x}^\nu(\tau,\sigma')\}=\frac{1}{T}\eta^{\mu,\nu}\delta(\sigma-\sigma'),
\]
\[
\{x^{\mu}(\tau,\sigma), {x}^\nu(\tau,\sigma')\}=\{\dot{x}^{\mu}(\tau,\sigma), \dot{x}^\nu(\tau,\sigma')\}=0.
\]

It is convenient to turn to light-cone coordinates on $C$ assuming light speed to be 1. They are $\zeta_{\pm}=\sigma\pm\tau$ and the flat metric $ds_C^2$ becomes
$ds_C^2=d\zeta_{+}d\zeta_{-}$. The metric components of $\eta_{\alpha\beta}$ in the light-cone coordinates are $\eta_{++}=\eta_{--}=0$ and $\eta_{+-}=\eta_{-+}=\frac{1}{2}$. The differential operators become
$\partial_{\pm}=\frac{1}{2}(\partial_{\sigma}\mp\partial_{\tau})$. 

The Virasoro generators are defined by
\[
L_m=\frac{1}{2}\int_0^{2\pi}T_{++}e^{im(\tau-\sigma)} d\sigma=\frac{1}{2}\sum\limits_{n\in\mathbb Z}\alpha_{m-n}\alpha_n,
\]
\[
\tilde{L}_m=\frac{1}{2}\int_0^{2\pi}T_{--}e^{im(\tau+\sigma)} d\sigma=\frac{1}{2}\sum\limits_{n\in\mathbb Z}\beta_{m-n}\beta_n,
\]
where $T_{++}$ and $T_{--}$ are the principal diagonal components of the energy-momentum tensor in the real light-cone coordinates and  $\alpha^{\mu}_0=\beta_0^{\mu}=\frac{1}{2\pi T}p^{\mu}$.
The commutator relation for $L_n$ given by the Poisson structure is
\[
\{L_m,L_n\}=i(n-m)L_{n+m},\quad \{\tilde{L}_m,\tilde{L}_n\}=i(n-m)\tilde{L}_{n+m},\quad \{L_m, \tilde{L}_n\}=0.
\]

The next useful operation is so-called Wick rotation when the non-physical time $\tau$ is replaced by the imaginary time $i\tau$. Then the light-cone coordinates change to $\zeta_{+}\to z$, $\zeta_{-}\to\bar{z}$, where $z=\sigma+i\tau$.
The differential operators become $\partial_{+}\to \partial_z$, $\partial_{-}\to\partial_{\bar{z}}$. Then the Polyakov action in the conformal gauge reads as
\[
S_P=-2T\int\limits_{C}\frac{dz\wedge d\bar{z}}{2i}\eta_{\mu\nu}\partial_zx^{\mu}\partial_{\bar{z}}x^{\nu},
\]
and the equation of motion is Laplacian $\partial^2_{z\bar{z}}x^{\mu}=0$. The energy-momentum tensor has the following components
\[
T_{zz}=\frac{1}{2}\eta_{\mu\nu}((x^{\mu})'-i\dot{x}^{\mu})((x^{\nu})'-i\dot{x}^{\nu})=T_{00}+2iT_{10},
\]
\[
T_{\bar{z}\bar{z}}=\frac{1}{2}\eta_{\mu\nu}((x^{\mu})'+i\dot{x}^{\mu})((x^{\nu})'+i\dot{x}^{\nu})=T_{00}-2iT_{10},
\]
and $T_{{z}\bar{z}}=T_{\bar{z}{z}}=0$. The Virasoro constraints in these coordinates are written as $T_{zz}=T_{\bar{z}\bar{z}}=0$.

The invariance of $S_P$ with respect to an infinitesimal reparametrization $\sigma^{\alpha}\to \tilde{\sigma}^{\alpha}+\varepsilon^{\alpha}(\sigma)$ implies
the conservation law for the energy-momentum tensor 
$\nabla^{\alpha}T_{\alpha\beta}=0$ which in $z$-coordinates becomes 
\[
\partial_{\bar{z}}T_{zz}+\partial_zT_{\bar{z}{z}}=\partial_zT_{\bar{z}\bar{z}}+\partial_{\bar{z}}T_{{z}\bar{z}}=0,
\]
and making use of $T_{{z}\bar{z}}=T_{\bar{z}{z}}=0$, we conclude that $T_{zz}$ is analytic and $T_{\bar{z}\bar{z}}$ is antianalytic components of $T$.

Then in complex coordinates
\[\partial_{z}x^{\mu}=\frac{1}{\sqrt{4\pi T}}\sum\limits_{n\in\mathbb Z}\beta_n^{\mu}e^{-inz},
\]
\[\partial_{\bar{z}}x^{\mu}=\frac{1}{\sqrt{4\pi T}}\sum\limits_{n\in\mathbb Z}\alpha_n^{\mu}e^{-in\bar{z}}.
\]

The Wick rotation leads to the complex coordinates and we conclude that in complex coordinates the Virasoro generators $L_n$ are
the coefficients of the formal Laurent series
\[
T_{zz}=\sum\limits_{n\in\mathbb Z}\frac{L_n}{z^{n+2}},
\]
and the commutator relations for $L_n$ in these coordinates are $$\{L_m,L_n\}=i(n-m)L_{n+m},$$ so span$\{L_n\}$ forms the Witt algebra.

\subsection{Canonical quantization}
In classical mechanics the motion is completely defined by the position $x$ and momentum $p$, which are the elements of a symplectic manifold. 
They specify the state of the classical mechanical system.
There is a specific function, the Hamiltonian function $H$, that defines
the motion by the equations $\dot{x}=\nabla_pH$, $\dot{p}=-\nabla_xH$. The symplectic structure (also known as the canonical structure) defines the Poisson bracket by $\{x,p\}=1$. 

The main goal of the classical mechanics is to find the trajectories of motion of particles, which are the solutions to a Hamiltonian system. The task of quantum  mechanics differs because
we have no trajectories any longer and a particle is found at a point of spacetime with some probability (complex in general). The entire information
about the quantum system is given by the spectrum and the corresponding collection of eigenfunctions for a quantum  operator. 
The initial task of the first (canonical) quantization is to find operators $X$ and $P$ corresponding to the dynamical variables $x$ and $p$, which act over an infinite dimensional Hilbert space of smooth functions, quantum states. The next step is to find quantum analogue of the Poisson bracket and the
operator, the Hamiltonian, which defines the total energy of a system, or in other words, an analogue of the Hamiltonian function. Finally, the states of the quantum system can be labelled by the eigenvalues of the Hamiltonian.

Let us start with the definition of the operators $X$ and $P$. In quantum mechanics we replace $x\to \cdot x\equiv X$, $p\to -i\hbar\nabla_x\equiv P$. Acting on quantum states they result in $X\psi=x\psi$, $P\psi=-i\hbar\nabla_x\psi$. There exist many reasons of this replacement. Let us give a
toy reason related to the wave function, which appears in the harmonic oscillator  with the phase $\varphi=kx-\omega t$, where $\omega$ is the frequency and $k$ is the coefficient of displacement (Hooke's law). The Planck constant is the proportionality constant between energy ($E$) of a photon and the frequency of its associated electromagnetic wave: $E=\hbar \omega$, $\hbar=6.62606896(33) \times 10^{-34}/2\pi\,J\,s$ (Joule$\times$seconds). The value $\hbar k$ is the momentum of the displacement in a given direction $x$. Then the expression $\hbar d\varphi=\hbar kdx-\hbar \omega dt$ can be interpreted as $\hbar d\varphi=pdx-Hdt=dS$, where $S$ is the action. The real wave function is $\cos \varphi$, and the complex wave function is $\psi=\exp(\frac{i}{\hbar}S)$ with the amplitude~1.
We consider the simplest case in which the amplitude is constant=1. Generally, it can be some non-negative function. 

Let us consider the action $S$, i.e., the time integral of the Lagrangian, on the extremal trajectories, the solutions to the Euler-Lagrange equation.
Then the result is a function $S(t_0,t, x_0, x)$, that depends on the initial and final time $t_0$ and $t$, and on the initial and final point of the trajectory
$x_0$ and $x$. Differentiating with respect to the final point $x$ gives $p=\nabla_x S$, where the momentum $p$ is the solution of the corresponding Hamiltonian system  $\dot{x}=\nabla_pH$, $\dot{p}=-\nabla_xH$ taken at the final time.
So we immediately obtain $-i\hbar\nabla\Psi=p\Psi$ and the momenta represent  eigenvalues of the operator $-i\hbar\nabla$. Thus, the quantum commutator is $[X,P]\psi=(XP-PX)\psi=(x(i\hbar)\nabla-i\hbar \nabla x)\psi=i\hbar\psi$, or $[X,P]=i\hbar$. The quantum Poisson brackets are to preserve the classical properties in which $\{x,p\}=1$, therefore we come to the conclusion
\[
\{X,P\}_{{\rm PB}}=\frac{[X,P]_{{\rm quantum}}}{i\hbar}.
\]

Moreover, in the canonical quantization the fields are replaced by Hermitian operators. Rescaling our bosonic string, such that $\hbar\to 1$, $T\to 1$, we have
\[
\{x^{\mu}(\tau,\sigma), \dot{x}^\nu(\tau,\sigma')\}=-i\eta^{\mu,\nu}\delta(\sigma-\sigma'),
\]
\[
\{x^{\mu}(\tau,\sigma), {x}^\nu(\tau,\sigma')\}=\{\dot{x}^{\mu}(\tau,\sigma), \dot{x}^\nu(\tau,\sigma')\}=0.
\]

When the canonical quantization procedure is applied in the frames of a given quantum field theory, the classical field variable becomes a quantum operator, which acts on a quantum state  to increase or decrease the number of particles by one. For a boson there are two operators: the boson's creation operator $\mathfrak b^{\dag}$ and the boson's annihilation operator $\mathfrak b$ (commonly known as `ladder operators') for the system with one degree of freedom, or $\mathfrak b^{\dag}_k$ and $\mathfrak b_k$ for fields, $k\in \mathbb Z$.  Each operator creates or annihilates particles in a certain state $\psi$. 
The next step in quantization is establishing the normal ordering operation between creators and annihilators following the Heisenberg uncertainty
principle expressed in above Poisson bracket. 
The commutation relations for creators and annihhilators are $\{\mathfrak b^{\dag},\mathfrak b^{\dag}\}=0$, $\{\mathfrak b,\mathfrak b\}=0$, and $\{\mathfrak b,\mathfrak b^{\dag}\}=1$. The latter can be rewritten as $\mathfrak b\mathfrak b^{\dag}=\mathfrak b^{\dag}\mathfrak b+1$. A product of creation and annihilation operators is said to be in the {\it normal order}  when all creation operators are to the left of all annihilation operators in the product. The process of putting a product into normal order is called normal ordering. The normal ordering operation is denoted by $:\mathfrak a \mathfrak b:$. In the case of two boson's operators $:\mathfrak b^{\dag} \mathfrak b:=\mathfrak b^{\dag}\mathfrak b$ but
$:\mathfrak b\mathfrak b^{\dag}:=\mathfrak b^{\dag}\mathfrak b$ and $\mathfrak b\mathfrak b^{\dag}=:\mathfrak b\mathfrak b^{\dag}:+1$.

We define the Virasoro generators in the quantum system by introducing normal ordering for positively indexed oscillators $\alpha_n$, $\beta_n$ as annihilators and for  the negatively indexed ones as creators:
\[
L_m=\frac{1}{2}\sum\limits_{n\in\mathbb Z}:\alpha_{m-n}\alpha_n:
\]
\[
\tilde{L}_m=\frac{1}{2}\sum\limits_{n\in\mathbb Z}:\beta_{m-n}\beta_n:
\]
where $m\neq 0$. The only operator $L_0$ produces an anomaly because it is already in normal order,
\[
L_0=\frac{1}{2}\alpha_0\alpha_0+\frac{1}{2}\sum\limits_{n\in\mathbb Z'}:\alpha_{-n}\alpha_n:=\frac{1}{2}\alpha_0\alpha_0+\frac{1}{2}\sum\limits_{n\in\mathbb Z'}\alpha_{-n}\alpha_n.
\]
In order to keep the commutation law for oscillators one must  subtract from $L_0$ some constant $c$ when taking commutator of symmetric operators $L_n$ and $L_{-n}$.
Taking into account quantum brackets one arrives at the Virasoro commutation relation
\begin{equation}\label{Vir}
\{L_m,L_n\}_{{\rm Vir}}=(n-m)L_{n+m}+\frac{c}{12}n(n^2-1)\delta_{n,-m},
\end{equation}
where the constant $c$ is called the central charge and in this particular case it coincides with the dimension $D$.

The Virasoro constraints can not be written as $L_n|\psi\rangle=0$ for all $n$, because 
\[
\langle \psi | \{L_{n}, L_{-n}\} | \psi\rangle=2n\langle \psi | L_{0} | \psi\rangle+\frac{c}{12}n(n^2-1)\langle \psi | \psi\rangle\neq 0.
\]
So the correct Virasoro constraints are imposed by  $L_n|\psi\rangle=0$ for $n>0$ and $(L_0-c)|\psi\rangle=0$. Here $\langle \psi|$ and $|\psi\rangle$  are the standard Dirac notations of {\it bra} and $\it ket$ vectors of states. Our system is 
 in the physical state $|\psi\rangle$ and $c=D$.
 
Early string theory proposed by Yoichiro Nambu and others in 1970 was only  bosonic. Pierre Ramond, Andr\'e Neveu, and John Schwarz completed the theory by inventing  fermionic strings to accompany the bosonic ones. 

\section{KdV and Virasoro algebra}

The KdV equation appeared in a paper by Diederik Johannes Korteweg and his student Gustav de Vries \cite{KdV} in Philosophical Magazine, 1895, and originally described the solitary wave discovered by the Scottish engineer John Scott Russell about half a century earlier. Later on, it became the condition on the potential of the spectral stability of the Schr\"odinger operator, eigenvalues of which represent energy quantization for bound systems. The existence of quantized energy levels is verified experimentally by observation of the energy emitted or absorbed when the system makes a transition from one level to another. 

\subsection{Hamiltonian dynamics and integrability}
In order to speak on integrability of KdV let us introduce this notion briefly for finite dimensional Hamiltonian systems which  will be used in the sequel. 
 There exists a vast amount
of modern literature dedicated to different approaches and
definitions of {\it integrable systems} (see, e.g., \cite{Arnold,
Babelon, Bolsinov, Zakharov}).

The classical definition of an {\it integrable system} in
the sense of Liouville applied to a Hamiltonian system says, that if we can
find  independent conserved integrals which are pairwise involutory
(have vanishing Poisson brackets), this system is partially or completely integrable depending on the number of pairwise involutory integrals
(see e.g., \cite{Arnold, Babelon, Bolsinov}). That is
each first integral allows us to reduce the order of the system not
just by one, but by two. We give now the precise definitions.

Let $(N,\omega)$ be a symplectic manifold with the symplectic 2-form $\omega$ and $H$ be a $C^{\infty}$-function defined on $N$. Then we write $\overrightarrow{H}$ to denote the {\it Hamiltonian vector field} associated to~$H$. By definition $\overrightarrow{H}$ is a vector field on $N$, such that $$\omega(\overrightarrow{V},\overrightarrow{H})=dH(\overrightarrow{V}),$$ for every smooth vector field $\overrightarrow{V}$ on $N$. In this case the function $H$ is called the {\it Hamiltonian function}. If $H,K\in C^{\infty}(N)$, then the Poisson bracket $\{H,K\}$ is the directional derivative of $K$ in the direction of $\overrightarrow{H}$, i.~e. $$\{H,K\}=dK(\overrightarrow{H})=\omega(\overrightarrow{H},\overrightarrow{K}).$$ The Jacoby identity and the Leibniz property show that the map $H\to \overrightarrow{H}$ is a Lie algebra homomorphism $$(C^{\infty}(N),\{\cdot,\cdot\})\ \ \text{to}\ \ (V^{\infty}(N),[\cdot,\cdot]),$$ where $V^{\infty}(N)$ denotes the set of $C^{\infty}$-vector fields on $N$. An equation of the form \begin{equation}
\dot{x}(t)=\overrightarrow{H}(x(t)),\label{Hamilton11}
\end{equation}
is called the Hamiltonian system. 

Let us consider the particular case of $n$-dimensional complex vector space $\mathbb C^n$. In this case the cotangent bundle $T^*(\mathbb C^n)$ for $\mathbb C^n$, that is, isomorphic to $\mathbb C^{2n}$, has a natural symplectic structure. Relatively to the coordinate chart $$T^*(\mathbb C^n)=(z_1,\ldots,z_n,\bar\psi_1,\ldots,\psi_n),$$ we have the formulas $$\omega=\sum_{k=1}^{n} dz^k \wedge d\bar\psi_k,$$ $$ \overrightarrow{H}=\sum_{k=1}^{n}\frac{\partial H}{\partial z^k}\frac{\partial}{\partial\bar\psi_k}-\frac{\partial H}{\partial\bar\psi_k}\frac{\partial}{\partial z^k},\qquad \{H,K\}=\sum_{k=1}^{n}\frac{\partial H}{\partial z^k}\frac{\partial K}{\partial\bar\psi_k}-\frac{\partial H}{\partial\bar\psi_k}\frac{\partial K}{\partial z^k}.$$
Then the system~\eqref{Hamilton11} can be written in the form
\begin{equation}
\dot{z}^k(t)=\frac{\partial H}{\partial \overline{\psi}_k},\quad
\dot{\overline{\psi}}_k(t)=-\frac{\partial H}{\partial z^k},\quad
k=1,2\dots,n.\label{Hamilton21}
\end{equation} and, using the Poisson brackets, we get 
\begin{equation}
\dot{z}^k(t)=\{z^k,H,\},\quad
\dot{\overline{\psi}}_k(t)=\{\overline{\psi}_k, H\},\quad
k=1,2\dots,n.\label{Hamilton31}
\end{equation} The latter system has $n$ degrees of freedom. A smooth function $\Phi$ satisfying $\{H,\Phi\}=0$ is called the {\it first integral} of the Hamiltonian system. In particular, $\{H,H\}=0$, and the Hamiltonian function $H$ is a first
integral of the system (\ref{Hamilton11}). If the system
(\ref{Hamilton31}) has $n$ functionally independent first integrals
$\Phi_1,\dots,\Phi_n$, which are pairwise involutory, that is 
$\{\Phi_k,\Phi_j\}=0$, $k,j=1,\dots,n$, then the system is called {\it
completely integrable} in the sense of Liouville. The function $H$
is included in the set of the first integrals. The classical theorem
of Liouville and Arnold \cite{Arnold} gives a  description
of the motion generated by the completely integrable system
(\ref{Hamilton31}). It states that such a system admits action-angle
coordinates on a connected regular compact invariant manifold.

\subsection{From KdV to Virasoro}

Let us consider the phase coordinates (field variables) defined on the spacetime $S^1\times \mathbb R$,  which are from the Sobolev space $H^{\alpha}(S^1)$, $\alpha\geq -1$,   regarding to $x\in S^1$ for each fixed $t\in \mathbb R$, i.e., real valued distributions on the unit circle  $u(e^{ix},t)$ . Let us simplify $u\to u(x,t)$, where the new $u$ becomes a $2\pi$ periodic smooth in $x$ function.
Observables are  $C^1$ functionals on $H^{\alpha}(S^1)$.

The famous  KdV equation is  $u_t=6uu'+u'''$ on the unit circle can be viewed as an integrable Hamiltonian system of infinite dimensions.
Several Poisson structures can be defined on the space of observables. One was proposed by Gardner~\cite{Gardner71}, and independently, by Zaharov and Faddeev~\cite{ZaharovFaddeev}, see others in, e.g.,~\cite{Enriques, Gorsky, Magri}.

Let us consider two functionals $F(u)$ and $G(u)$, where $u\in H^{\alpha}(S^1)$ with the vanishing mean value. Expanding $u$ into the Fourier series we obtain
\[
u(x)=\sum\limits_{n\in \mathbb Z'}u_ne^{inx}, 
\]
where $u_{-n}=\bar{u}_n$, and $\mathbb Z'=\mathbb Z\setminus \{0\}$. Then let us consider the functionals $F$ and $G$ as functions $F=F(q_1,q_2,\dots, p_1,p_2\dots)$ and $G=G(q_1,q_2,\dots, p_1,p_2\dots)$ with respect $q_n=u_n/n$ and $p_n=\bar{u}_{-n}$, $n\in \mathbb Z'$.
The canonical symplectic form $dq\wedge d\bar{p}$ implies the Poisson structure
\[
\{F,G\}=\frac{1}{2\pi}\sum\limits_{n=1}^{\infty}\left(\frac{\partial F}{\partial q_n}\frac{\partial G}{\partial \bar{p}_n}-\frac{\partial F}{\partial \bar{p}_n}\frac{\partial G}{\partial q_n}\right)=\frac{1}{2\pi}\sum\limits_{n\in \mathbb Z'}n\frac{\partial F}{\partial u_n}\frac{\partial G}{\partial u_{-n}}.
\]
Observe that $u_0$ can be included now in the latter sum.

Take into account a useful formula for variational derivatives. If $$F=\int_{0}^{2\pi}f(u,u',u'',\dots)dx,$$ then
\[
\frac{\delta F}{\delta u}=\frac{\partial f}{\partial u}-\frac{d}{dx}\frac{\partial f}{\partial u'}+\frac{d^2}{dx^2}\frac{\partial f}{\partial u''}-\dots
\]
Hence, if the function $u$ depends on a parameter $\alpha$, then integrating by parts we conclude that
\[
\frac{d F}{d\alpha}=\int_{0}^{2\pi}\frac{\delta F}{\delta u}\frac{\p u}{\p \alpha}dx.
\]
In particular,
\[
\frac{\partial F}{\partial u_n}=\int_{0}^{2\pi}\frac{\delta F}{\delta u}e^{inx}dx,\quad \frac{\delta F}{\delta u}=\frac{1}{2\pi}\sum\limits_{n\in \mathbb Z'}\frac{\partial F}{\partial u_n}e^{inx}=\frac{1}{2\pi}\sum\limits_{n\in \mathbb Z'}\frac{\partial F}{\partial u_{-n}}e^{-inx}.
\]
Substituting this in the formula for the Poisson bracket we obtain
\[
\{F,G\}=\frac{1}{4\pi^2}\int\limits_0^{2\pi}\sum\limits_{n\in \mathbb Z'}n\frac{\partial F}{\partial u_n}\frac{\partial G}{\partial u_{-n}}dx=\frac{1}{4\pi^2}\int\limits_0^{2\pi}\sum\limits_{n\in \mathbb Z'}\frac{\partial F}{\partial u_n}e^{inx}\frac{\partial G}{\partial u_{-n}}ne^{-inx}dx
\]
\[
=\frac{1}{4\pi^2}\int\limits_0^{2\pi}\sum\limits_{n\in \mathbb Z'}\frac{\partial F}{\partial u_n}e^{inx}\sum\limits_{k\in \mathbb Z'}\frac{\partial G}{\partial u_{-k}}ke^{-ikx}dx=i\int\limits_0^{2\pi}\frac{\delta F}{\delta u}\frac{d}{d x}\frac{\delta G}{\delta u}dx,
\]
which is true for all functionals of the above integral form.

If we choose the Hamiltonian function in the form $H= -i\int_0^{2\pi} (\frac{1}{2}(u'^2)+u^3) dx$, then the evolution equation $\dot{u}=\{u,H\}$ admits the form
\[
\dot{u}=i\frac{d}{d x}\frac{\delta H}{\delta u}=6uu'+u'''.
\]

One of the most important features of this equation is that it possesses an infinite number of conserved quantities (first integrals) $I_k[u]$, e.g., 
\[
I_{-1}=\int_0^{2\pi} u dx,\,\,\, I_{0}=\int_0^{2\pi} u^2 dx, \,\,\, I_{1}=\int_0^{2\pi} (\frac{1}{2}(u'^2)+u^3) dx,\dots\]

\[ \dots, I=\int_0^{2\pi} \mbox{polynomial $(\frac{d}{dx}, \cdot u)dx$}.
\]
 which are all in involution. One observes the relation of this fact to the Liouville integrability
 for Hamiltonian systems. However, the proof of complete integrability is rather difficult task, which was fulfilled in \cite{Gardner71, ZaharovFaddeev}.

If we consider the conserved quantities as Hamiltonians, then we obtain a hierarchy  constructed as
 \[
 \dot{u}=\{u, -iI_n\}\equiv\frac{d}{dx}\frac{\delta I_n}{\delta u},
 \]
which is called the {\it KdV hierarchy}.

The above Poisson structure considered on the phase space 
formally can be rewritten (modulo a constant factor) as
 \[
[u(x),u(y)]=\gamma \delta'(x-y), \quad \gamma>0.
 \]
It is degenerate because the center  is one-dimensional and the admissible element 
 \[
P=\int_0^{2\pi}u(x)dx.
 \]
commutes with all observables. Fixing $P$ we get a submanifold which is symplectic.
The Poisson structure on the phase space gives the Lie structure on the space of observables.

Hamiltonian itself is an observable, and for example, consideration of
 \[
 H=\frac{1}{2\gamma}\int_0^{2\pi} u^2(x,t)dx
 \]
 (corresponding to $I_0$) gives a trivial evolution equation of motion $\dot{u}=\{H,u\}=-u'$, $u=u(x-t)$.
   The infinite
number of evolution equations generated by all integrals is the above KdV hierarchy.
 The Miura transformation $s=u^2+u'$ reduces the modified KdV equation $\dot u= u'''+u^2u'$ to the
usual KdV and leads to the Hamiltonian
 \[
 H=\frac{1}{2\gamma}\int_0^{2\pi}( u^2+u')dx.
 \]  The Poisson brackets for $s$ satisfy the relation
 \[
 \{s(x), s(y)\}=\gamma(2(s(x)+s(y))+\delta'(x-y)+\delta'''(x-y),
 \] 
which is called {\it Magri brackets}~\cite{Magri}.

Fourier coefficients of $s(x)$ are given as
 \[
 L_n(s)=\frac{1}{2\pi}\int_0^{2\pi} e^{inx}s(x) dx.
 \]   
The corresponding Lie brackets are $\{L_m,L_n\}_{{\rm Vir}}$
 where $c$ is the central charge, $c=\frac{6\pi}{\gamma}$ (J.-L.Gervais \cite{Gervais}), or taking into account quantum effects $c=1+6(\frac{\pi}{\gamma}+\frac{\gamma}{\pi}+2)$, see  \cite{Faddeev} (one can observe here the famous gap in CFT for real $c\in (1,25)$).

\subsection{From Virasoro to KdV}

Let us define
\[
u=\frac{6}{c}\sum_{n\in \mathbb Z}L_ne^{-inx}-\frac{1}{4}
\]   
 Then, using $\delta(x)=\frac{1}{2\pi}\sum_{n\in \mathbb Z}e^{inx}$ and the Virasoro commutation relation~\eqref{Vir}, we obtain
 \[
 \{u(x), u(y)\}=\frac{6\pi}{c}(-\delta'''(x-y)+4u(x)\delta'(x-y)+2u'\delta(x-y)).
 \]

Taking $I_0=\frac{1}{2}\int_{0}^{2\pi}u^2dx$, we obtain
\[
\dot{u}=\frac{c}{6\pi}\{u, I_0\}=u'''+6uu'.
\]

KdV as a non-linear PDE is related to the classical and quantum field theories in which the infinite number of degrees of freedom follows
from the infinite number of degrees of freedom for the initial conditions. 
So it is not surprising to see  relations between the Virasoro algebra and KdV. As we shall show in forthcoming sections, problems of completely different nature (the L\"owner-Kufarev evolution) but also of the infinite number of degrees of freedom, lead to a rigid algebraic structure
given by the Virasoro algebra.

\section{Realization on the unit circle}

Mathematically, the Virasoro algebra appeared first as a central  extension of the Lie algebra of smooth vector fields $\phi\frac{d}{d\theta}$ on the unit circle   $S^1$ (see \cite{GelfandFuchs}).  Let us denote the Lie group of $C^{\infty}$ sense preserving
diffeomorphisms of the unit circle $S^1$  by $\Diff S^1$. Each element of $\Diff S^1$ is represented as
$z=e^{i\alpha(\theta)}$ with a monotone increasing $C^{\infty}$
real-valued function $\alpha(\theta)$, such that
$\alpha(\theta+2\pi)=\alpha(\theta)+2\pi$.
The space of smooth vector fields on a differentiable manifold $S^1$ forms a Lie algebra, where the Lie bracket is defined to be the commutator of vector fields. The relation of this Lie algebra to $\Diff S^1$ is subtile. The Lie algebra to $\Diff S^1$ can be associated with the left-invariant vector fields  $\Vect S^1$. 
But the exponential map, which is the same as the exponential map from the tangent space at the origin, is not even locally a homeomorphism. 
 The infinitesimal action of $\Vect S^1$ is $\theta\to\theta+\varepsilon \phi(\theta)$. To $\phi$ we associate the vector field $\phi\frac{d}{d\theta}$, and the Lie
brackets are given by
\begin{equation}\label{Lie1}
[\phi_1,\phi_2]={\phi}_1{\phi}'_2-{\phi}_2{\phi}'_1.
\end{equation}
 The  Virasoro algebra is the unique (up to isomorphism) non-trivial central extension of $\Vect\, S^1$ by $\mathbb R$ given by the {\it Gelfand-Fuchs cocycle}~\cite{GelfandFuchs}.

\subsection{Canonical identification}
As an infinite dimensional Lie-Fr\'echet group, $\Diff S^1$ undergoes certain irregular behaviour, in particular, the exponential map from $\Vect S^1$ is not a local homeomorphism. The entire necessary background of unitary representations of $\Diff S^1$ is found in the study of Kirillov's
homogeneous K\"ahlerian manifold $\Diff S^1/S^1$.
We  deal with the analytic representation of \linebreak
$\Diff S^1/S^1$. Let $\Sb$ stands for the whole class of univalent functions $f$ in the unit disk $U$ normalized by $f(z)=z(1+\sum_{n=1}^{\infty}c_nz^n)$ about the origin.
By $\tilde{\Sb}$ we denote the class of functions from $\Sb$ smooth ($C^{\infty}$) on the boundary $S^1$ of $U$.
Given a map $f\in \tilde\Sb$ we construct the adjoint univalent
meromorphic map 
\[
g(z)=d_1z+d_0+\frac{d_{-1}}{z}+\dots,
\] 
defined in the exterior $U^*=\{z:\,|z|>1\}$ of $U$, and such that  $\hat{\mathbb C}\setminus\overline{f(U)}=g(U^*)$. Both functions are extendable onto $S^1$. This conformal welding gives the identification of the homogeneous manifold $\Diff S^1/S^1$ with the space $\tilde{\Sb}$:  $\tilde{\Sb}\ni f\leftrightarrow f^{-1}\circ g|_{S^1}\in\Diff S^1/S^1$, or with the  smooth contours $\Gamma=f(S^1)$ that enclose univalent domains $\Omega$ of conformal radius 1 with respect to the origin and such that $\infty\not\in \Omega$, $0\in\Omega$, see \cite{Airault}, \cite{KY1}. 
Being quasicircles, the smooth contours allow us to embed $\Diff S^1/S^1$ into the  universal Teichm\"uller space making use of the above conformal welding, see \cite{TakTeo}. Coefficients of the univalent functions from $\tilde{\Sb}$ are the natural coordinates on the Teichm\"uller space. So one can construct complexification
of $\Vect S^1$ and further projection of the holomorphic part to the set $\mathcal M\subset \mathbb C^{\mathbb N}$ which is the projective limit of the coefficient bodies $\mathcal M=\lim_{n\leftarrow \infty}\mathcal M_n$, where
\begin{equation}
\mathcal M_n=\{(c_1,\dots,c_n):\,\,f\in \tilde{\Sb}\}.\label{Mn}
\end{equation}
This construction relates the K\"ahler structure of both manifolds.
The holomorphic Virasoro generators can then be realized by the first order differential operators
\[
L_j=\partial_j+\sum\limits_{k=1}^{\infty}(k+1)c_{k}\partial_{j+k},\quad j\in \mathbb N,
\]
in terms of the affine coordinates of $\mathcal M$, acting over the set of holomorphic functions, where $\partial_{k}=\partial/\partial{c_k}$. We explain the details in the next subsection.

\subsection{Complexification}
Let us introduce local coordinates on the manifold \linebreak $\mathcal M=\Diff S^1/S^1$ in the concordance with the local coordinates on the space  $\tilde{\Sb}$ of univalent functions smooth on the boundary. Observe that $\mathcal M$ is a real infinite-dimensional manifold, whereas $\tilde{\Sb}$
is a complex manifold. We are aimed at a complexification of $T\mathcal M$ which admits a holomorphic projection to $T\tilde{\Sb}$, where as usual,
$\Vect_0 S^1=\Vect\, S^1/\const$ is a module over the ring of smooth functions, which is associated with the tangent bundle $T\mathcal M$. 

Two operations are to be considered: {\it complexification}, {\it conjugation}, and {\it almost complex structure}. Given a real vector space $V$ the complexification $V_{\mathbb C}$ is defined as the tensor product with the complex numbers $V\otimes_{\mathbb R}\mathbb C$, that often is written as $V_{\mathbb C}=V\oplus iV$. The subscript $\mathbb R$ indicates that we take the real tensor product, we omit it in the sequel. Elements of $V_{\mathbb C}$ are of the form $v\otimes z$.
In addition, the  vector space $V_{\mathbb C}$ is a complex vector space that follows by defining multiplication by complex numbers,
$\alpha(v\otimes z)=v\otimes \alpha z$ for complex $\alpha$ and $z$ and $v\in V$. The space $V$ is naturally embedded into $V\otimes \mathbb C$ by identifying $V$ with $V\otimes 1$.
 Conjugation
is defined by introducing a canonical conjugation map on $V_{\mathbb C}$ as $\overline{v\otimes z}=v\otimes \bar z$.

An almost {\it complex structure} $J$ on $V$ is a linear transformation $J: V\to V$ such that $J^2=-I$. 
It can be extended by linearity to the complex structure $J$ on $V_{\mathbb C}$ by $J(v\otimes z)=J(v)\otimes z$. Observe that 
\[
\overline{J(v\otimes z)}=\overline{Jv\otimes z}=Jv\otimes \bar{z}=J(v\otimes \bar{z})=J(\overline{v\otimes z}).
\]

Eigenvalues of extended $J$ are $\pm i$, and there are two eigenspaces $V^{(1,0)}$ and $V^{(0,1)}$ corresponding to them given by projecting $\frac{1}{2}(1\mp iJ)v$. $V_{\mathbb C}$ is decomposed into the direct sum $V_{\mathbb C}=V^{(1,0)}\oplus V^{(0,1)}$, where $V^{(1,0)}=\{v\otimes 1- J(v)\otimes i\big| v\in V \}$ and $V^{(0,1)}=\{v\otimes 1+ J(v)\otimes i\big| v\in V \}$. In the case of existence of such a complex structure $J$, the vector spaces $V^{(1,0)}$ and $V^{(0,1)}$ give complex coordinates on $V$.

An almost complex structure on $\Vect_0 S^1$ may be defined as follows (see \cite{Airault}).
We  identify $\Vect_0 S^1$  with the functions with vanishing mean value over~$S^1$. It gives
\[
\phi(\theta)=\sum\limits_{n=1}^{\infty}a_n\cos\,n\theta+b_n\sin\,n\theta.
\]
Let us define an almost complex structure by the operator
\[
J(\phi)(\theta)=\sum\limits_{n=1}^{\infty}-a_n\sin\,n\theta+b_n\cos\,n\theta.
\]
Then $J^2=-id$. On $\Vect_0S^1\otimes \mathbb C$, the operator $J$ diagonalizes and we have the identification
\[
\Vect_0S^1\ni \phi\leftrightarrow v:=\frac{1}{2}(\phi-iJ(\phi))=\sum\limits_{n=1}^{\infty}(a_n-ib_n)e^{in\theta}\in (\Vect_0S^1\otimes \mathbb C)^{(1,0)},
\]
and the latter extends into the unit disk as a holomorphic function.

The Kirillov infinitesimal action \cite{Kir2} of $\Vect_0 S^1$ on $\tilde \Sb$ is given by
a variational formula due to Schaeffer and Spencer \cite[page 32]{Schaeffer} which lifts the actions from the Lie algebra $\Vect_0
S^1$ onto $\tilde\Sb$. Let $f\in\tilde{\Sb}$ and let
$\phi(e^{i\theta}):=\phi(\theta)\in \Vect_0 S^1$ be a $C^{\infty}$ real-valued function in
$\theta\in(0,2\pi]$. The infinitesimal action
$\theta \mapsto \theta+\varepsilon \phi(e^{i\theta})$ yields a variation of the univalent function $f^*(z)= f+\varepsilon\,\delta_{v}f(z)+o(\epsilon)$, where
\begin{equation}
\delta_{v}f(z)=\frac{f^2(z)}{2\pi
}\int\limits_{S^1}\left(\frac{wf'(w)}{f(w)}\right)^2\frac{v(w)dw}{w(f(w)-f(z))} ,\label{var}
\end{equation}
and $\phi\leftrightarrow v$ by the above identification.
 Kirillov and Yuriev \cite{KY1}, \cite{KY2} (see also \cite{Airault})  established
that the variations $\delta_{\phi}f(\zeta)$ are closed with respect to
the commutator (\ref{Lie1}), and the induced Lie algebra is the same as $\Vect_0
S^1$.  The Schaeffer-Spencer operator is linear.

Treating $T\mathcal M$ as a real vector space, the operator $\delta_{\phi}$ transfers the complex structure $J$ from $\Vect_0 S^1$ to $T\mathcal M$ by $J(\delta_{\phi}):=\delta_{J(\phi)}$. By abuse of notation,  we denote the new complex structure on $T\mathcal M$  by the same character $J$. Then it splits the complexification $T\mathcal M_{\mathbb C}$ into two eigenspaces $T\mathcal M_{\mathbb C}=T\mathcal M^{(1,0)}\oplus T\mathcal M^{(0,1)}$. Therefore,  $\delta_{v}=\delta_{\phi-iJ(\phi)}:=\delta_{\phi}-iJ(\delta_{\phi})\in T\mathcal M^{(1,0)}$. 
Observe that $2z\partial_z=-i\partial_{\theta}$ on the unit circle $z=e^{i\theta}$, and $L_k=z^{k+1}\partial/\partial z=-\frac{1}{2}ie^{ik\theta}\partial/\partial \theta$ on $S^1$. Let us take the basis
of $\Vect_0 S^1\otimes \mathbb C$ in the form $\nu_k=-ie^{ik\theta}$ in order to keep the index of vector fields the same as for $L_k$. Then, the commutator satisfies the Witt relation $\{\nu_m, \nu_n\}=(n-m)\nu_{n+m}$. Taking  elements $\nu_k=-iw^k$, $|w|=1$ in the integrand of (\ref{var}) we calculate the residue in (\ref{var}) and obtain so called Kirillov operators
\[
L_j[f](z)=\delta_{\nu_j}f(z)=z^{j+1}f'(z), \quad j=1,2,\dots,
\]
which are the  holomorphic coordinates on $T\mathcal M^{(1,0)}$.
In terms of the affine coordinates in $\mathcal M$ we get the Kirillov operators as
\[
L_j=\partial_j+\sum\limits_{k=1}^{\infty}(k+1)c_{k}\partial_{j+k},
\]
where $\partial_k=\partial/\partial c_k$. They satisfy the Witt commutation relation $$\{L_m,L_n\}=(n-m)L_{n+m}.$$
For $k=0$ we obtain the operator $L_0$, which corresponds to the constant vectors from $\Vect \,S^1$, $L_{0}[f](z)=zf'(z)-f(z)$.
The elements of  the Fourier basis $-ie^{-i\theta k}$ with negative indices (corresponding to $T\mathcal M^{(0,1)}$) are extended into $U$ by
$-iz^{-k}$. Substituting them in (\ref{var}) we get very complex formulas for $L_{-k}$, which functionally depend on $L_k$ (see \cite{Airault}, \cite{Kir2}) and  might play the role of conjugates to $L_k$. The first two operators are calculated as
\begin{eqnarray*}
L_{-1}[f](z)&=&f'(z)-2c_1 f(z)-1,\\
L_{-2}[f](z)&=&\frac{f'(z)}{z}-\frac{1}{f(z)}-3c_1+(c_1^2-4c_2)f(z),
\end{eqnarray*}
see \cite{KY2}.

This procedure gives a nice links between representations of the Virasoro algebra and the theory of univalent functions.
The L\"owner-Kufarev equations proved to be a powerful tool to work with univalent functions (the famous Biberbach conjecture was proved \cite{Branges} using L\"owner method). In the following section we show how L\"owner-Kufarev equations can be used in a representation of the Virasoro algebra. In particular, we identify $T\mathcal M^{(1,0)}$ with  $T\mathcal M$, equipped with its natural complex structure given by coefficients of univalent functions, by means the L\"owner-Kufarev PDE.

\section{L\"owner-Kufarev Equations}

A time-parameter family $\Omega(t)$ of simply connected hyperbolic univalent domains forms a {\it L\"owner subordination chain}  in the complex plane $\mathbb
C$, for $0\leq t< \tau$ (where $\tau$ may be $\infty$), if
$\Omega(t)\varsubsetneq \Omega(s)$, whenever $t<s$.
We
suppose that the origin is an interior point of the Carath\'eodory kernel of
$\{\Omega(t)\}_{t=0}^{\tau}$.  

A L\"owner subordination chain $\Omega(t)$ is described by a time-dependent family of conformal maps $z=f(\zeta,t)$
from the unit disk $U=\{\zeta:\,|\zeta|<1\}$ onto $\Omega(t)$, normalized by $f(\zeta,t)=a_1(t)\zeta+a_2(t)\zeta^2+\dots$,
$a_1(t)>0$, $\dot{a}_1(t)>0$. After L\"owner's 1923 seminal  paper \cite{Loewner} a fundamental contribution to
the theory of L\"owner chains was made by Pommerenke \cite{Pommerenke1, Pommerenke2} who
described governing evolution equations in partial and ordinary derivatives, known now as
the L\"owner-Kufarev equations due to Kufarev's work \cite{Kufarev}.

One can normalize the growth of 
evolution of a subordination chain by the conformal radius of
$\Omega(t)$ with respect to the origin by  $a_1(t)=e^t$.

 L\"owner \cite{Loewner}  studied a
time-parameter semigroup of conformal one-slit maps of the unit disk $U$ arriving
then at an evolution equation called after him. His main
achievement was an infinitesimal description of the semi-flow of
such maps by the Schwarz kernel that led him to the L\"owner
equation. This crucial result was then generalized in several
ways (see \cite{Pommerenke2} and the references therein).

We say that the  function $p$ is from the Carath\'eodory class if it is analytic in $U$, normalized as $p(\zeta)=1+p_1\zeta+p_2\zeta^2+\dots,\quad
\zeta\in U,$ and such that $\re p(\zeta)>0$ in~$U$.
Pommerenke \cite{Pommerenke1, Pommerenke2} proved that given a subordination
chain of domains $\Omega(t)$ defined for $t\in [0,\tau)$, there exists
a function $p(\zeta,t)$, measurable in $t\in [0,\tau)$ for any fixed $z\in U$, and from the Carath\'eodory class for almost all   $t\in [0,\tau)$, such that 
the conformal mapping $f:U\to \Omega(t)$ solves the equation
\begin{equation}
\frac{\partial f(\zeta,t)}{\partial t}=\zeta\frac{\partial
f(\zeta,t)}{\partial \zeta}p(\zeta,t),\label{LK}
\end{equation}
for $\zeta\in U$ and for almost all $t\in [0,\tau)$.   The
equation (\ref{LK}) is called the L\"owner-Kufarev equation due to
two seminal papers: by L\"owner \cite{Loewner} who considered the case when
\begin{equation}
p(\zeta,t)=\frac{e^{iu(t)}+\zeta}{e^{iu(t)}-\zeta},\label{yadro}
\end{equation}
where $u(t)$ is a continuous function regarding to $t\in [0,\tau)$,
 and by Kufarev \cite{Kufarev} 
who proved differentiability of $f$ in $t$ for all $\zeta$ from the kernel of $\{\Omega(t)\}$ in the case of general $p$ in the Carath\'eodory class.

 Let us consider a reverse process. We are given an
initial domain $\Omega(0)\equiv \Omega_0$ (and therefore, the
initial mapping $f(\zeta,0)\equiv f_0(\zeta)$), and a function
$p(\zeta,t)$ of positive real part normalized by $p(\zeta,t)=1+p_1\zeta+\dots$. Let us solve the equation (\ref{LK})
and ask ourselves, whether the solution $f(\zeta,t)$ defines a subordination
chain of simply connected univalent domains $f(U,t)$. The initial condition
$f(\zeta,0)=f_0(\zeta)$ is not given on the characteristics of the
partial differential equation (\ref{LK}), hence the solution exists
and is unique but not necessarily univalent. Assuming $s$ as a parameter along the characteristics
we have $$ \frac{dt}{ds}=1,\quad \frac{d\zeta}{ds}=-\zeta
p(\zeta,t), \quad \frac{df}{ds}=0,$$ with the initial conditions
$t(0)=0$, $\zeta(0)=z$, $f(\zeta,0)=f_0(\zeta)$, where $z$ is in
$U$.  Obviously, $t=s$. Observe that the domain of $\zeta$ is the entire unit disk. However, the solutions to
the second equation of the characteristic system range within the unit disk but do not fill it. 
Therefore, introducing another letter $w$ (in order to distinguish the function $w(z,t)$  from the variable $\zeta$) we arrive at the Cauchy problem for the  L\"owner-Kufarev
equation in ordinary derivatives
\begin{equation}
\frac{dw}{dt}=-wp(w,t),\label{LKord}
\end{equation}
 for a function $\zeta=w(z,t)$
with the initial condition $w(z,0)=z$. The equation (\ref{LKord}) is a non-trivial  characteristic
equation for (\ref{LK}). Unfortunately, this approach requires the
extension of $f_0(w^{-1}(\zeta,t))$ into the whole $U$ ($w^{-1}$ means the inverse function) because the solution to
(\ref{LK}) is the function $f(\zeta,t)$  given as
$f_0(w^{-1}(\zeta,t))$, where $\zeta=w(z,s)$ is a solution of the
initial value problem for the characteristic equation (\ref{LKord})
that maps $U$ into $U$. Therefore, the solution of the initial
value problem for the equation (\ref{LK}) may be non-univalent.

 Solutions to the
equation (\ref{LKord}) are holomorphic univalent functions
$w(z,t)=e^{-t}z+a_2(t)z^2+\dots$ in the unit disk that map $U$  into
itself. Every function $f$ from the class $\Sb$ can be
represented by the limit
\begin{equation}
f(z)=\lim\limits_{t\to\infty}e^t w(z,t),\label{limit}
\end{equation}
where $w(z,t)$ is a solution to \eqref{LKord}  with some  function $p(z,t)$ of positive real part for almost
all $t\geq 0$ (see \cite[pages 159--163]{Pommerenke2}). Each function
$p(z,t)$ generates a unique function from the class $\Sb$. The
reciprocal statement is not true. In general, a function $f\in \Sb$
can be obtained using different functions $p(\cdot,t)$. 

Now we are ready to formulate the condition of  univalence of the solution to the equation (\ref{LK}), which
can be obtained by combination of known results of \cite{Pommerenke2}.

\begin{theorem}\label{ThProkhVas}{\rm \cite{Pommerenke2, ProkhVas}} Given  a function
$p(\zeta,t)$ of positive real part normalized by $p(\zeta,t)=1+p_1\zeta+\dots$, the solution to   the equation (\ref{LK})
is unique, analytic and univalent with respect to $\zeta$ for almost all $t\geq 0$, if and only if, the initial condition
$f_0(\zeta)$ is taken in the form \eqref{limit}, where the function $w(\zeta,t)$ is the solution to the equation \eqref{LKord}
with the same driving function $p$.
\end{theorem}

Recently, we started to look at L\"owner-Kufarev equations from the point of view of  motion in the space of univalent functions where Hamiltonian and Lagrangian formalisms play a central role (see, \cite{Vasiliev}). Some connections with the Virasoro algebra were also observed in \cite{MPV, Vasiliev}. The present paper generalizes these attempts and gives their closed form.  The main conclusion is that the L\"owner-Kufarev equations are naturally linked to the holomorphic part of the Virasoro algebra.
Taking holomorphic Virasoro generators $L_n$ as a basis of the tangent space to the  coefficient body for univalent functions at a fixed point, we see that the driving function in the L\"owner-Kufarev theory generates generalized moments for motions within
the space of univalent functions. Its norm represents the energy of this motion. The holomorphic Virasoro generators in their co-tangent form will become  conserved quantities of the L\"owner-Kufarev ODE. The L\"owner-Kufarev PDE becomes a transition formula from the affine basis to Kirillov's basis of the holomorphic part of the complexified tangent space to $\mathcal M$ at any point.  Finally, we propose to study an alternate L\"owner-Kufarev evolution
instead of subordination.

\section{Witt algebra and the classical L\"owner-Kufarev equations}

In the following subsections we reveal the structural role of the Witt algebra as a background of the classical L\"owner-Kufarev contour evolution. As we see further,
the conformal anomaly and the Virasoro algebra appear as a quantum or stochastic effect in SLE.

\subsection{L\"owner-Kufarev ODE}
Let us consider the functions
\[
w(z,t)=e^{-t}z\left(1+\sum\limits_{n=1}^{\infty}c_n(t)z^n\right),
\]
satisfying the L\"owner-Kufarev ODE
\begin{equation}\label{LKord1}
\frac{dw}{dt}=-wp(w,t),
\end{equation}
with the initial condition $w(z,0)=z$, and with the function
$p(z,t)=1+p_1(t)z+\dots$ which is holomorphic in $U$ and measurable with respect to
$t\in [0,\infty)$, such that $\re p>0$ in $U$. The function
$w(z,t)$ is univalent and maps $U$ into $U$. 

\begin{lemma}
 Let the function $w(z,t)$ be a solution to the Cauchy problem for the equation
\eqref{LKord1}
with the initial condition $w(z,0)=z$. If the driving function $p(\cdot,t)$, being from the Carath\'eodory class for almost all $t\geq 0$, is  $C^{\infty}$ smooth in the closure $\hat{U}$ of the unit disk $U$ and summable with respect to $t$, then the boundaries of
the domains $B(t)=w(U,t)\subset U$ are smooth for all $t$.
\end{lemma}
\begin{proof}
Observe that the continuous and differentiable dependence of
 the solution of a differential equation $\dot{x}=F(t,x)$ on the initial condition $x(0)=x_0$ is a classical problem. One can refer, e.g., to \cite{Volpato}
in order to assure that summability of $F(\cdot, x)$ regarding to $t$ for each fixed $x$ and  continuous differentiability ($C^1$ with respect to
$x$ for almost all $t$) imply that the solution $x(t,x_0)$ exists, is unique, and is $C^1$ with respect to $x_0$. In our case, the solution to  \eqref{LKord1}
exists, is unique and analytic in $U$, and, moreover, $C^1$ on its boundary $S^1$. Let us differentiate \eqref{LKord1} inside the unit disk $U$ with respect to $z$ and write
\[
\log w' =-\int\limits_{0}^{t}(p(w(z,\tau),\tau)+w(z,\tau)p'(w(z,\tau),\tau))d\tau,
\]
choosing the branch of the logarithm such as $\log w'(0,t)=-t$.
This equality is extendable onto $S^1$ because the right-hand side is, and therefore, $w'$ is $C^1$ and $w$ is $C^2$ on $S^1$. We continue analogously and write the formula
\[
w''=-w'\int\limits_{0}^{t}(2w'(z,\tau)p'(w(z,\tau),\tau)
+w(z,\tau)w'(z,\tau)p''(w(z,\tau),\tau))d\tau,
\]
which guarantees that $w$ is $C^3$ on $S^1$. Finally, we come to the conclusion that $w$ is $C^\infty$ on $S^1$.
\end{proof}

Let 
$f(z,t)$ denote $e^tw(z,t)$. The limit $\lim_{t\to\infty}f(z,t)$ is known
 \cite{Pommerenke2} to be a representation of all univalent functions. 
 
  Let the driving term $p(z,t)$ in the L\"owner-Kufarev ODE be from the Carath\'eodory class for almost all $t\geq 0$,   $C^{\infty}$ smooth in $\hat{U}$, and summable with respect to $t$.
 Then the domains $\Omega(t)=w(U,t)$  have  smooth boundary $\partial
 \Omega(t)$. So the L\"owner equation can be extended onto the
 closed unit disk $\hat U=U\cup S^1$.
 
 Consider the Hamiltonian given by
 \begin{equation}\label{Ham3}
 H=\int\limits_{z\in S^1}f(z,t)(1-p(e^{-t}f(z,t),t))\bar \psi(z,t)\frac{dz}{iz},
 \end{equation}
on the unit circle $z\in S^1$, where $\psi(z,t)$ is a formal series 
$$
\psi(z,t)=\sum_{n=-k}^{\infty}\psi_nz^{n},
$$
defined about the unit circle $S^1$ for any $k\geq 0$.
The Poisson structure on the space $(f, \bar\psi)$ is given by the canonical brackets
\[
\{P, Q\}=\frac{\delta P}{\delta f}\frac{\delta Q}{\delta \bar \psi}-\frac{\delta P}{\delta \bar \psi}\frac{\delta Q}{\delta f},
\]
or in  coordinate form (only $\psi_n$ for $n\ge 1$ are independent co-vectors corresponding to the tangent vectors $\partial_n$ with respect to the canonical Hermitean product for analytic functions)
\[
\{p, q\}=\sum_{n=1}^{\infty}\frac{\partial p}{\partial c_n}\frac{\partial q}{\partial \bar \psi_n}-\frac{\partial p}{\partial \bar \psi_n}\frac{\partial q}{\partial c_n}.
\]
 Here
\[
P(t)= \int\limits_{z\in S^1}p(z,t)\frac{dz}{iz},\quad Q(t)= \int\limits_{z\in S^1}q(z,t)\frac{dz}{iz}.
\]
The Hamiltonian system becomes
\begin{equation}\label{sys1}
\frac{d f(z,t)}{dt}=f(1-p(e^{-t}f,t))=\frac{\delta H}{\delta \overline{\psi}}=\{f,H\},
\end{equation}
for the position coordinates and
\begin{equation}\label{sys2}
\frac{d\bar \psi}{dt}=-(1-p(e^{-t}f,t)-e^{-t}fp'(e^{-t}f,t))\bar
\psi=\frac{-\delta H}{\delta f}=\{\overline{\psi},H\},
\end{equation}
for the momenta, where $\frac{\delta}{\delta f}$ and $\frac{\delta}{\delta \overline{\psi}}$ are the variational derivatives. So the phase coordinates $(f,\bar{\psi})$ play the role of the canonical Hamiltonian pair.

The coefficients $c_n$ are the complex local coordinates on $\mathcal M$, so in these coordinates we have
\begin{eqnarray}
\dot{c}_n & = &c_n-\frac{e^t}{2\pi i}\int\limits_{S^1}w(z,t)p(w(z,t),t)\frac{dz}{z^{n+2}}, \nonumber\\ 
&=&-\frac{1}{2\pi i}\int\limits_{S^1}\sum\limits_{k=1}^ne^{-kt}(e^tw)^{k+1}p_k\frac{dz}{z^{n+2}},\quad n\geq 1.\nonumber
\end{eqnarray}
Let us fix some $n$ and project the infinite dimensional Hamiltonian system on an $n$-dimensional $\mathcal M_n$.
Momenta in coordinates
form an adjoint vector \linebreak $\bar\psi(t)=(\bar\psi_1(t),\dots, \bar\psi_n(t))^T$,
with complex-valued coordinates $\psi_1,\dots,\psi_n$ for any fixed $n$. The dynamical equations for momenta governed by the Hamiltonian function \eqref{Ham3} are
\begin{equation*}
\dot{\bar{\psi}}_j=-\bar{\psi}_j+\frac{1}{2\pi i}\sum\limits_{k=1}^n\bar{\psi}_k\int\limits_{S^1}(p+wp')\frac{dz}{z^{k-j+1}},\quad j= 1,\dots, n-1,\label{psi1}
\end{equation*}
and 
\begin{equation}
\dot{\bar{\psi}}_n=0.
\end{equation}
In particular,
\begin{eqnarray*}
\dot{c}_1 & = & -e^{-t}p_1,\\
\dot{c}_2 & = & -2e^{-t}p_1c_1-e^{-2t}p_2,\\
\dot{c}_3 & = & -e^{-t}p_1(2c_2+c_1^2)-3e^{-2t}p_2c_1-e^{-3t}p_3,\\
\dots& & \dots
\end{eqnarray*}
for $n=3$ we have
\begin{eqnarray*}
\dot{\bar{\psi}}_1 & = & 2e^{-t}p_1\bar{\psi}_2+(2e^{-t}p_1c_1+3e^{-2t}p_2)\bar{\psi}_3,\\
\dot{\bar{\psi}}_2 & = & 2e^{-t}p_1\bar{\psi}_3,\\
\dot{\bar{\psi}}_3 & = & 0.
\end{eqnarray*}

 Let us set the function  $L(z):=f'(z,t)\bar\psi(z,t)$. Let $(L(z))_{<0}$ mean the part of the Laurent series for $L(z)$ with negative powers of $z$,
$$
(L(z))_{<0}=(\bar\psi_1+2c_1\bar\psi_2+3c_2\bar\psi_3+\dots)\frac{1}{z}+(\bar\psi_2+2c_1\bar\psi_3+\dots)\frac{1}{z^2}+\dots=\sum\limits_{k=1}^{\infty}\frac{L_k}{z^{k}}.
$$
 Then, the functions $L(z)$ and $(L(z))_{< 0}$ are time-independent
for all $z\in S^1$.

It is easily seen that, passing from the cotangent vectors
$\bar\psi_k$ to the tangent vectors $\partial_k$, the
coefficients $L_k$ of $(L(z))_{<0}$ defined on the tangent bundle  $T\mathcal M^{(1,0)}$ are exactly the Kirillov vector fields $L_k$.  Corresponding co-vector fields $L_k$ are
conserved by the L\"owner-Kufarev ODE because $\dot L_k=\{L_k, H\}=0$. The above Poisson structure coincides with  that given by the Witt brackets introduced for $L_k$ previously. For finite-dimensional grades
this result was obtained in \cite{MPV}.

Let us formulate the result as a theorem.

\begin{theorem} Let the driving term $p(z,t)$ in the L\"owner-Kufarev ODE be from the Carath\'eodory class for almost all $t\geq 0$,   $C^{\infty}$ smooth in $\hat{U}$, and summable with respect to $t$. Then the Kirillov co-vector fields are the conserved quantities for the Hamiltonian system
(\ref{sys1}--\ref{sys2}) generated by the L\"owner-Kufarev ODE.
\end{theorem}

\begin{remark}
Another way to construct a Hamiltonian system could be based on the symplectic structure given by the K\"ahlerian form on $\Diff S^1/S^1$. However, there is no explicit expression for such  form in terms of  functions $f\in \tilde\Sb$. Moreover, there must be a Hamiltonian formulation
in which the L\"owner-Kufarev equation becomes an evolution equation. This remains an open problem.
\end{remark}

\begin{remark}
At a first glance the situation with an ODE with  a parameter is quite simple. Indeed, if we solve an equation of type $\dot{f}(t, e^{i\theta})=F(f(t, e^{i\theta}),t)$, then fixing $\theta$ we have an integral  of motion $C=I(f(t,\cdot),t)=\const$. Then, releasing $\theta$, we have $C(e^{i \theta})=I(f(t,e^{i\theta}),t)$. Expanding $C(e^{i\theta})$ into the Fourier series, we obtain an infinite number of conserved quantities, but they do not manifest an
infinite number of degrees of freedom that govern the motion as in the field theory where the governing equations are PDE. In our case, we have not only
one trajectory fixing the initial condition but a pensil of trajectories because our equation has an infinite number of control parameters, the Taylor coefficients of the function $p(z,t)$, which form a bounded non-linear set of admissible controls. Therefore, we operate with sections of the tangent and co-tangent bundles  to the inifinite dimensional manifold $\mathcal M$ instead of vector fields along one trajectory as in usual ODE.
\end{remark}

\begin{remark}
No linear combinations $L^*_k$ of $L_1,\dots,L_n,\dots$ allows us to reduce the system of $\{L_k\}$ to a new system of involutory $\{L_k^*\}$ in order
to claim the Liouville integrability of our system. Observe that the coefficients in these linear combinations must be constants to keep conservation laws. 
\end{remark}

\subsection{Construction of $L_0$ and $L_{-n}$}
Consider again the generating function $L(z)=f'(z,t)\bar\psi(z,t)$ and the `non-negative' part  $(L(z))_{\ge 0}$ of the Laurent series for $L(z)$,
 $$
(L(z))_{\ge 0}= (\bar\psi_0+2c_1\bar\psi_1+3c_2\bar\psi_2+\dots)+ (\bar\psi_{-1}+2c_1\bar\psi_0+3c_2\bar\psi_1+\dots)z+\dots
$$
$$
=\sum\limits_{k=0}^{\infty}\mathcal L_{-k}z^{k}.
$$
All $\mathcal L_{-k}$ are conserved by the construction. Define $\bar\psi_0^*=-\sum_{n=1}^{\infty}c_k\bar\psi_k$, and
\[
L_0=\mathcal L_0-(\bar\psi_0-\bar\psi^*_0).
\]
The operator $L_0$ acts on the class $\Sb$ by  $L_0[f](z)=zf'(z)-f(z)$.
 Next define  $L_{-1}=\mathcal L_{-1}-(\bar\psi_{-1}-\bar\psi_{-1}^*)-2c_1(\bar\psi_0-\bar\psi_0^*)$, where $\bar\psi_{-1}^*=0$. Then,
\[
L_{-1}[f](z)=f'(z)-2c_1 f(z)-1
\]
Finally, $$L_{-2}=\mathcal L_{-2}-(\bar\psi_{-2}-\bar\psi_{-2}^*)-2c_1(\bar\psi_{-1}-\bar\psi_{-1}^*)-3{c_2}(\bar\psi_0-\bar\psi_0^*).$$
We choose $\bar\psi_{-2}^*=(c_3-3c_1c_2+c_1^3)\bar\psi_1+\dots$, so that
\[
\bar\psi_{-2}^*[f](z)=\frac{1}{z}-\frac{1}{f(z)}-c_1-(c_2-c_1^2)f(z),
\]
and
\[
L_{-2}[f](z)=\frac{f'(z)}{z}-\frac{1}{f(z)}-3c_1+(c_1^2-4c_2)f(z).
\]
An important fact is that 
\[
L_{0}=c_1 \bar\psi_1+2c_2 \bar\psi_2+\dots,
\]
\[
L_{-1}=(3c_2-2c_1^2)\bar\psi_1+\dots,
\]
\[
L_{-2}=(5c_3-6c_1c2+2c_1^3) \bar\psi_1+\dots,
\]
are linear with respect to $ \bar\psi_k$, $k\geq 1$, and therefore, are co-vectors. Equivalently $$L_{0,-1,-2}[f](z)=\mbox{function}(c_1,c_2,\dots)z^2+\dots, \quad z^k=\frac{\partial f}{\partial c_{k-1}}.$$
All other co-vectors we construct by our Poisson brackets as 
\[
L_{-n}=\frac{1}{n-2}\{L_{-n+1},L_{-1}\}=\frac{1}{n-4}\{L_{-n+2},L_{-2}\}.
\]
The form of the Poisson brackets guarantees us that all $L_{-n}$ are linear with respect to $\bar\psi_1,\bar\psi_2,\dots$ and span
the anti-holomorphic part of the co-tangent bundle ${T^{(0,1)}}^*\mathcal M$.

Let us summarize the above in the following conclusion.
We considered a non-linear contour dynamics given by the L\"owner-Kufarev equation. It turned out to be underlined by an algebraic structure, namely, by the Witt algebra spanned by the Virasoro generators $L_n$, $n\in\mathbb Z$.
\begin{itemize}
\item[$\bullet$]   $L_n$, $n = 1,2,\dots$ are the holomorphic Virasoro generators. They span a holomorphic vector bundle over the space of univalent functions $\Sb$, smooth on the boundary. In their co-vector form, $L_n$ are conserved by the L\"owner-Kufarev evolution.

\item[$\bullet$] $L_0$ is the central element.

\item[$\bullet$]  $L_{-n}$, $n = 1,2,\dots$ are the antiholomorphic Virasoro generators. They span the antiholomorphic vector bundle.  In their co-vector form, $L_{-n}$ contain a conserved part and we give an iterative method to obtain all of them based on the Poisson structure of the L\"owner-Kufarev evolution.
\end{itemize}

\subsection{L\"owner-Kufarev PDE}

 The L\"owner equation in partial derivatives is
$${ \dot{w}(\zeta,t)=\zeta w'(\zeta,t)p(\zeta,t)}, \quad \re p(\zeta,t)>0, \quad |\zeta|<1.$$
with some initial condition $w(z,0)=f_0(z)$.
Let us consider the one-parameter family of functions
$f(z,t)=e^{-t}w(z,t)=z(1+\sum_{n=1}^{\infty}c_n(t)z^n)$, $f(z,0)=f_0(z)$ as a $C^1$ path
in $\tilde\Sb$. At the identity $id$ we have that $T_{id}\tilde\Sb=T_{id}\mathcal M^{(1,0)}=T_{id}\mathcal M$. A path in the coefficient body $\mathcal M$ in the neighbourhood of the identity is
 $(c_1(t),\dots, c_n(t),\dots)$ with the velocity vector
$\dot c_1\partial_1+\dots+\dot c_n\partial_n+\dots \in T_{id}\mathcal M$. 

Taking the Virasoro generators $\{L_k\}$, $k\geq 1$, as an algebraic basis in $T\mathcal M_{id}^{(1,0)}$ we wish the velocity vector written in this new basis to be
\begin{equation}
{ \dot c_1\partial_1+\dots+\dot c_n\partial_n+\dots=u_1L_1+\dots
u_n L_n+\dots}.\label{e1}
\end{equation}
We compare (\ref{e1})  with the L\"owner-Kufarev equation
\begin{equation} { \dot f=\dot c_1\partial_1+\dots+\dot
c_n\partial_n+\dots=zf'p(z,t)- f=L_0+u_1L_1+\dots u_n L_n+\dots},\label{e2}
\end{equation}
where $p(z,t)=1+u_1z+\dots+u_nz^n+\dots$, and $L_0f=zf'-f$. In view of  similarity between these two expressions \eqref{e1} and \eqref{e2}, we notice that 
\begin{itemize}
\item a new term $L_0$ appears in the L\"owner-Kufarev equation;
\item the function $p(z,t)$ with positive real part corresponds to subordination, whereas for generic trajectories it may have  real part of arbitrary sign. We call this an {\it alternate L\"owner-Kufarev evolution};
\item the vector $L_0$  corresponds exactly to the rotation: $$e^{i\varepsilon}f(e^{-i\varepsilon}z)=f(z)-i\varepsilon(zf'(z)-f(z))+o(\varepsilon).$$  
\end{itemize}

 Let us consider the set $\tilde{\Sb}_0$ of non-normalized smooth univalent functions
of the form $F(z,t)=a_0(t)z+a_1(t)z^2+\dots$, with a tangent vector $\dot a_0\partial_0+\dots+\dot a_n\partial_n+\dots$, where $\partial_k=\partial/\partial a_k$, $k=0,1,2, \dots$. Our aim is to define two distributions of co-dimension 1 for the tangent bundle $T\tilde{\Sb}_0$,  that form the tangent bundle
  $T\tilde{\Sb}$. This will be realized by means of formulas
  \eqref{e1} and \eqref{e2}.   Notice that  $\partial_k F=z^{k+1}$. Setting $L_k(F):=z^{k+1}F'$ we get
\[
\dot F=\dot a_0\partial_0+\dots+\dot
a_n\partial_n+\dots=zf'p(z,t)=u_0L_0+u_1L_1+\dots u_n L_n+\dots,
\]
where $p(z,t)=u_0+u_1z+\dots+u_nz^n+\dots$. This alternate L\"owner-Kufarev equation represents
recalculation of the tangent vector in the new basis
\[
{ \dot a_0\partial_0+\dots+\dot a_n\partial_n+\dots=u_0L_0+\dots
u_n L_n+\dots},
\]
where $L_k=a_0\partial_k+2a_1\partial_{k+1}+\dots$.

Let us present the distributions. We start with $F\in \tilde{\Sb}_0$, then we define $f\in \tilde{\Sb}$. The necessary distribution is the map
$$\tilde{\Sb}_0\ni F\to T_f\tilde{\Sb}\hookrightarrow T_F\tilde{\Sb}_0.$$

The analytic form of the first distribution is the following factorization
 $f_1(z,t)=\frac{1}{a_0}F(z,t)=z+\frac{a_1}{a_0}z^2+\dots$, so that
\begin{equation}
 \dot f_1=zf_1'p(z,t)-\frac{\dot a_0}{a_0}f_1,\label{F1}
 \end{equation}
 where $u_0=\frac{\dot a_0}{a_0}$. Then we obtain $$ { \dot c_1\partial_1+\dots+\dot c_n\partial_n+\dots=\hat L_0+u_1\hat L_1+ \dots +u_n \hat L_n+\dots}$$
where $\hat L_0f_1=u_0(zf_1'-f_1)$, $\hat L_kf_1=z^{k+1}f_1'$, $c_k=\frac{a_k}{a_0}$,
$\partial_k=\frac{\partial}{\partial c_k}$.  In
particular, $a_0=e^t$ implies the L\"owner-Kufarev equation for arbitrary sign of $\re p$.

The analytic form of the second distribution becomes
 $f_2(z,t)=F(\frac{1}{a_0}z,t)=z+\frac{a_1}{a^2_0}z^2+\dots$, so that
\begin{equation}
 \dot f_2=zf_2'p(\frac{z}{a_0},t)- \frac{\dot a_0}{a_0} zf'_2,\label{F2}
 \end{equation}
 where again $u_0=\frac{\dot a_0}{a_0}$.
In the coefficient form we get
$$ { \dot c_1\partial_1+\dots+\dot c_n\partial_n+\dots=u_1\tilde L_1+ \dots +u_n \tilde L_n+\dots}$$
where  $\tilde L_kf_2=z^{k+1}f_2'$, $c_k=\frac{a_k}{a^{k+1}_0}$,
$\partial_k=\frac{\partial}{\partial c_k}$.

Observe that the equation  \eqref{F2} gives another identification of $T\mathcal M^{(1,0)}$ with  $T\mathcal M$.

Finally, let us make an explicit calculation of $\hat{L}_0$, which for $a_0=e^t$ we continue to denote by $L_0$. Using Kirilov's basis $L_1,L_2,\dots$ as a linear combination we write
$$L_0=\sum_{m=1}^{\infty}\Pi_mL_m.$$ The coefficients $\Pi_m$ are polynomials, which can be
obtained  using the following recurrent formulas
$$K_1=0,\ \ K_m=-\sum_{j=1}^{m-1}j(m-j+1)c_{m-j}c_j,\qquad
\Pi_m=mc_m+\sum_{j=1}^{m}K_{m-j+1}P_{j-1},$$ where $P_k$ are
polynomials
\begin{equation}\label{pol1}
P_0=1,\quad P_1=-2c_1,\quad P_2=4c^2_1-3c_2,\quad P_k=-\sum_{j=1}^{k}(j+1)c_jP_{k-j},
\end{equation} 

Let us summarize the above considerations in the following theorem.

\begin{theorem}
The L\"owner-Kufarev PDE \eqref{F1} gives the distribution of co-dimension 1 inside the tangent bundle $T\tilde{\Sb}_0$ of non-normalized smooth univalent functions  $\tilde{\Sb}_0$,  that forms the tangent bundle
  $T\tilde{\Sb}$. 
  
The equation \eqref{F2} gives another distribution, and moreover,  
 it makes the explicit correspondence between the natural  complex structure 
of   $T\tilde\Sb$ and the complex structure of $T\mathcal M^{(1,0)}$ at each point $f\in \tilde\Sb$.
\end{theorem}

One of the reason to consider the alternate L\"owner-Kufarev PDE is the regularized canonical Brownian motion on smooth Jordan curves.
For all Sobolev metrics $H^{\frac{3}{2}+\varepsilon}$, the classical theory of stochastic flows allows to construct Brownian motions on $C^1$ diffeomorphism group of $S^1$. The case 3/ 2 is critical. Malliavin \cite{Malliavin} constructed the canonical Brownian motion on the Lie algebra $\Vect S^1$ for the Sobolev norm $H^{3/2}$. Another construction was proposed in \cite{Fang}. Airault and Ren \cite{AiraultRen} proved that the infinitesimal version
of the Brownian flow is H\"older continuous with any exponent $\beta<1$.

The regularized canonical Brownian motion on $\Diff S^1$ is a stochastic flow on $S^1$ associated to the It{\^{o}} stochastic differential equation
\[
dg^r_{x,t}=d\zeta^r_{x,t}(g^r_{x,t}),
\]
\[
\zeta^r_{x,t}(\theta)=\sum_{n=1}^{\infty}\frac{r^{n}}{\sqrt{n^3-n}}(x_{2n}(t)\cos n\theta-x_{2n-1}(t)\sin n\theta),
\]
where $\{x_k\}$ is a sequence of independent real-valued Brownian motions and $r\in (0,1)$ and the series for $\zeta^r_{x,t}(\theta)$ is a Gaussian trigonometric series. Kunita's theory of stochastic flows asserts that
the mapping $\theta\to g^r_{x,t}(\theta)$ is a $C^{\infty}$ diffeomorphism and the limit $\lim\limits_{r\to 1^{-}}g^r_{x,t}=g_{x,t}$ exists uniformly
in $\theta$. The random homeomorphism $g_{x,t}$ is called {\it canonical Brownian motion} on $\Diff S^1$, see \cite{AiraultRen, Fang, Malliavin, RenZhang}. It was shown in \cite{AiraultRen, Fang}, that this random homeomorphism is H\"older continuous.

The canonical Brownian motion can be defined not only on $\Diff S^1$, but  also on the space of $C^{\infty}$-smooth Jordan curves by conformal welding. This leads to dynamics of random loops which are not subordinated.

\section{Elliptic operators over the coefficient body}

The Kirillov first order differential operators $L_k$ generate the elliptic operator $\sum |L_k|^2$. In this section we construct the geodesic equation and find geodesics with constant velocity coordinates in the field of this operator. In particular, we shall prove that the norm of the driving function in the L\"owner-Kufarev theory gives the minimal energy of the motion
in this field.

\subsection{Dynamics within the coefficient body}
This subsection is auxiliary.
Let us  recall the geometry of the coefficient body  $\mathcal M_n$ for finite $n$ with respect to   Kirillov's basis $L_k$, $k=1,\dots,n$ of the tangent bundle $T\mathcal M_n$. The affine coordinates are introduced by projecting
$$
\mathcal M\ni f=z\Big(1+\sum\limits_{k=1}^{\infty}c_kz^k\Big)\mapsto (c_1,\ldots, c_n)\in\mathcal M_n.
$$
The manifold $\mathcal M_n$ was studied actively in the middle of the last century, see e.g., \cite{Babenko, Schaeffer}.
We compile some important properties of $\mathcal M_n$ below:
\begin{itemize}

\item[(i)] $\mathcal M_n$ is homeomorphic to a $(2n-2)$-dimensional ball and its
boundary $\partial \mathcal M_n$ is homeomorphic to a $(2n-3)$-dimensional
sphere;

\item[(ii)] every point $x\in \partial \mathcal M_n$ corresponds to exactly one function $f\in
\Sb$ which is called a {\it boundary function} for $\mathcal M_n$;

\item[(iii)] boundary functions map the unit disk $U$ onto the complex plane
$\mathbb C$ minus piecewise analytic Jordan arcs forming a tree with
a root at infinity and having at most $n$ tips,

\item[(iv)] with the exception for a set of smaller dimension,
at every point $x\in \partial \mathcal M_n$ there exists a normal vector
satisfying the Lipschitz condition;

\item[(v)] there exists a connected open set $X_1$ on $\partial \mathcal M_n$,
such that the boundary $\partial \mathcal M_n$ is an analytic hypersurface at
every point of $X_1$. The points of $\partial \mathcal M_n$ corresponding to
the functions that give the extremum to a linear functional belong
to the closure of $X_1$.

\end{itemize}

Properties (ii) and (iii) imply that the functions from $\tilde \Sb$ deliver interior points
of $\mathcal M_n$. The Kirillov operators $L_j$ restricted onto $\mathcal M_n$
give truncated vector fields
$$
L_j=\partial_j+\sum\limits_{k=1}^{n-j}(k+1)c_k\partial_{j+k},
$$
which we, if it causes no confusion, continue denoting by $L_j$ in this section.
In~\cite{MPV} based on the L\"owner-Kufarev representation, we showed that these $L_j$ can be obtained from a partially integrable Hamiltonian system for the coefficients in which the first integrals coincide with $L_j$.

Let $c(t)=\big(c_1(t),\ldots,c_{n}(t)\big)$ be a smooth trajectory in $\mathcal M_n$; that is a $C^1$ map $c:[0,1]\to \mathcal M_n$. Then the velocity vector $\dot c(t)$ written in the affine basis as $\dot c(t)=\dot c_1(t)\p_1+\ldots+\dot c_{n}(t)\p_{n}$ can be also represented in the basis of vector fields $L_1,\ldots,L_{n}$ (compare with \eqref{F2}) as \begin{eqnarray}\label{eq4}
\dot c(t) & =\dot c_1(t)\p_1+\ldots+\dot c_{n}(t)\p_{n}\\                                             & = u_1L_1+u_2L_3+\ldots+u_{n}L_{n}, \nonumber                                                                       \end{eqnarray}
where the coefficients $u_k$ can be written in the recurrent form  as 
\begin{equation}\label{eq3}
u_1=\dot c_1,\qquad u_k=\dot c_k-\sum_{j=1}^{k-1}(j+1)\dot c_ju_{k-j}.
\end{equation} Expressing  $u_k$ in terms of $c_k$ and $\dot c_k$, we get
\begin{equation}\label{eq8}
u_k=\dot c_k+\sum_{j=1}^{k-1}P_{j}\dot c_{k-j}.
\end{equation}

One may notice that these polynomials are the first coefficients of the holomorphic function $1/f'(z)$, where $f\in \tilde\Sb$. In the infinite dimensional case this follows from the L\"owner-Kufarev equation
 \eqref{F2} with $a_0=e^t$. Kirillov's fields $L_k$ act over these polynomials as  
 $$
 L_kP_n=(n-2k-1)P_{n-k}\quad n\geq k\quad\text{and}\quad L_kP_n=0 \quad n< k.
 $$

\begin{proposition}
Let us give the conjugate to $\{L_1,\ldots,L_n\}$ basis of one-forms. We define
\begin{eqnarray}\label{eq2}
\omega_1 & = & dc_1,\nonumber \\
\omega_2 & = & dc_2-2c_1\omega_1,\nonumber\\
\ldots & \ldots & \ldots\ldots\ldots\ldots\ldots,\nonumber\\
\omega_n & = & dc_n-\sum_{j=1}^{n-1}(j+1)c_j\omega_{n-j}.
\end{eqnarray}
Then
$$
\omega_n(L_n)=1,\quad \omega_n(L_k)=0\ \ \text{if}\ \ k\neq n.
$$ 
\end{proposition}
\begin{proof}
If $k>n$, then the vector fields $L_k$ do not contain $\p_n$ for $k>n$. Since the form $\omega_n$ depends only on $dc_j$ with $j<n$, then $$\omega_n(L_k)=\p_n(L_k)-\sum_{j=1}^{n-1}(j+1)c_j\omega_{n-j}(L_k)=0\ \ \text{for}\ \ k>n>n-j.$$ If $n=k$, then $$\omega_n(L_n)=\p_n(L_n)-\sum_{j=1}^{n-1}(j+1)c_j\omega_{n-j}(L_n)=1+0\ \ \text{for}\ \ n>n-j.$$ To prove the case $k<n$ we apply the induction. Let us show for $L_1$. We have $$\omega_2(L_1)=dc_2(L_1)-2c_1(L_1)=2c_1-2c_1=0.$$ We suppose that $\omega_n(L_1)=0$. Then $$\omega_{n+1}(L_1)=dc_{n+1}(L_1)-\sum_{j=1}^{n}(j+1)c_j\omega_{n+1-j}(L_1)=(n+1)c_{n}-(n+1)c_n\omega_1(L_1)=0.$$ The same arguments work for $\omega_n(L_k)$ with $k<n$.
\end{proof}

In the affine basis the forms can be written making use of the
polynomials $P_n$. We observe that one-forms $\omega_k$ are defined
in a similar way as the coordinates $u_k$ with respect to
the Kirillov vector fields $L_k$. Thus, if we develop the recurrent
relations~\eqref{eq2} and collect the terms with $dc_n$ we get
$$\omega_k=dc_k+\sum\limits_{j=1}^{k-1}P_jdc_{k-j}.\quad
k=1,\ldots,n.$$

By the duality of tangent and co-tangent bundles the information about the motion is encoded by these one-forms.

\subsection{Hamiltonian equations}
There exists an Hermitian form on $T\mathcal M_n$, such that the system
$\{L_1,\ldots,L_n\}$ is orthonormal with respect to this form. The
operator $L=\sum |L_k|^2$ is  elliptic, and  we write the
Hamilton function $H(c,\bar c, \psi, \bar\psi)$ defined on the
co-tangent bundle, corresponding to the operator $L$ as $H(c,\bar c,
\psi, \bar\psi)=\sum_{k=1}^{n}|l_k|^2$, where
$$l_k=\bar \psi_k+\sum_{j=1}^{n-k}(j+1)c_j\bar \psi_{k+j}.$$ The
corresponding Hamiltonian system admits the form \begin{eqnarray*}
\dot c_1 & = & \frac{\p H}{\p\bar\psi_1}= \bar l_1 \\
\ldots & = & \ldots\ldots\ldots\ldots \\
\dot c_n & = & \frac{\p H}{\p\bar\psi_n}=\bar l_n+\sum_{j=1}^{n-1}(j+1) c_j\bar l_{n-j}\\
\dot{\bar\psi}_p & = & -\frac{\p H}{\p c_p}=-(p+1)\sum_{k=1}^{n-p} l_k\bar \psi_{k+p}\\
\ldots & = & \ldots\ldots\ldots\ldots \\
\dot{\bar\psi}_n & = & -\frac{\p H}{\p c_n}=0.
\end{eqnarray*} Let us observe that
\begin{equation}\label{eq6}\dot l_k=\sum_{j=1}^{n-k}(j-k)\bar l_jl_{j+k}.\end{equation}
Expressing $\bar l_k$ from the first $n$ Hamilton equations we get
\begin{equation}\label{eq7}\bar l_k=\dot c_k+\sum_{j=1}^{k-1}P_{j}\dot c_{k-j},\quad k=1,\ldots,n.\end{equation} We can decouple the Hamiltonian system making use of~\eqref{eq6} and~\eqref{eq7} which leads us to the following non-linear differential equations of the second order
$$\ddot c_k=\dot{\bar l}_k+\sum_{j=1}^{k-1}(j+1)c_j\dot{\bar  l}_{k-l}+\sum_{j=1}^{k-1}(j+1)
\dot c_j\bar l_{k-l},$$ where $\dot{ l}_k$ are expressed in terms of the
product of $\bar l_jl_{j+k}$ by~\eqref{eq6}, and the last products depend on
$P_j$, $\bar P_j$ and $\dot c$, $\dot{\bar  c}_j$ for the
corresponding indices $j$ by \eqref{eq7}. For example,
$$
\ddot c_1  = \dot{\bar  l}_1=\sum_{j=1}^{n-1}(j-1)\Big(\dot
c_j+\sum_{p=1}^{j-1}P_p \dot c_{j-p}\Big)\overline{\Big(\dot
c_{j+1}+\sum_{q=1}^{j}P_q\dot c_{j+1-q}\Big)}.
$$

Comparing~\eqref{eq7} and~\eqref{eq8}, we conclude that $\bar
l_k=u_k$ and $u_k$ satisfy the differential equations
\begin{equation}\label{eq9}\dot u_k=\sum_{j=1}^{n-k}(j-k)\bar u_ju_{j+k},\end{equation} 
on the solution of the Hamiltonian system.
Observe that any solution of~\eqref{eq9} has a velocity vector of constant length. It is easy to see from the following system
\begin{eqnarray}\label{skewsym}
\bar u_1\dot u_1 & = & 0 \bar u_1\bar u_1u_2+\bar u_1\bar u_2u_3+2\bar u_1\bar u_3u_4+3\bar u_1\bar u_4u_5+4\bar u_1\bar u_5u_6+\ldots,\nonumber\\
\bar u_2\dot u_2 & = & -1 \bar u_1\bar u_2u_3+ 0 \bar u_2\bar u_2u_4+1\bar u_2\bar u_3u_5+2\bar u_2\bar u_4u_6+\ldots,\nonumber\\
\bar u_3\dot u_3 & = & -2 \bar u_1\bar u_3u_4-1 \bar u_2\bar u_3u_5+0\bar u_3\bar u_3u_6+\ldots,\\
\bar u_4\dot u_4 & = & -3 \bar u_1\bar u_4u_5-2 \bar u_2\bar u_4u_6+\ldots,\nonumber\\
\bar u_5\dot u_5 & = & -4 \bar u_1\bar u_5u_6+\ldots,\nonumber\\
\bar u_6\dot u_6 & = & \ldots\nonumber
\end{eqnarray} Then, $$\frac{d|u|^2}{dt}=2\sum_{k=1}^{n}(\bar u_k\dot u_k+u_k\dot{\bar u}_k)=0,$$ for any $n$, thanks to the cut form of our vector fields and the skew symmetry of \eqref{skewsym}. The simplest solution may be deduced for constant driving terms  $u_k$, $k=1,\ldots,n$. The Hamiltonian system immediately gives the geodesic
\begin{eqnarray*}
c_1 & = & \bar u_1(0)s+c_1(0),\\
c_2 & = & \bar u_1^2(0)s^2+\bar u_2(0)s+c_2(0),\\
c_3 & = & 3\bar u_1(0)\big(\bar u_1^2(0)\frac{s^3}{3}+\bar u_2(0)\frac{s^2}{2}+c_2(0)\big)+2\bar u_2(0)
\big(\bar u_1(0)\frac{s^2}{2}+c_1(0)s\big)+\bar u_3(0)s+c_2(0),\\
\ldots & = & \ldots\ldots\ldots\ldots\ldots\ldots
\end{eqnarray*}
In general, $c_n$ becomes a polynomial of order $n$ with coefficients
that depend on the initial data $c(0)$ and on the initial velocities
$\bar u(0)$.

The Lagrangian $\mathcal L$ corresponding to the Hamiltonian function $H$  can
be defined by the Legendre transform as $$\mathcal L=(\dot c,\bar
\psi)-H=\sum_{k=1}^{n}\Big(\bar l_k\bar \psi_k+\bar
\psi_k\sum_{j=1}^{k-1}(j+1)c_j\bar
l_{k-j}\Big)-\frac{1}{2}\sum_{k=1}^{n}|l_k|^2.$$ Taking into account that 
\[
\bar \psi_k\dot c_k =\sum\limits_{j=1}^{k-1} (j+1)c_j\bar{\psi}_k\bar{l}_{k-j}+\bar{\psi}_k\bar{l}_k.
\]
Summing up over $k$, we obtain $(\dot c,\bar
\psi)=\sum_{k=1}^{n}l_k\bar l_k=\sum_{k=1}^{n}\bar u_ku_k$, that
gives us
$$\mathcal L(c,\dot c)=\frac{1}{2}\sum_{k=1}^{n}|u_k|^2.$$ 
All these considerations can be generalized for $n\to\infty$.
Thus, we conclude that the coefficients of the function $p(z,t)$ in the L\"owner-Kufarev PDE play the
role of generalized moments for the dynamics in $\mathcal M_n$ and $\mathcal M$ with respect to the Kirillov basis on the tangent bundle. Moreover,
the $L^2$-norm of the function $p$ on the circle $S^1$ is the energy of such motion.

\section{SLE and CFT}

In this section we review the connections between conformal field theory (CFT)  and  Schramm-L\"owner evolution (SLE) following, e.g., \cite{BB}, \cite{FriedrichWerner}). SLE (being, e.g., a continuous limit of CFT's archetypical Ising model at its critical point) gives an approach to CFT which emphasizes CFT's roots in statistical physics. 

SLE$_{\varkappa}$ is a $\varkappa$-parameter family of covariant processes describing the evolution of random sets called the SLE$_{\varkappa}$ hulls. For different values of $\varkappa$ these sets can be either a simple fractal curve $\varkappa\in [0,4]$, or a self-touching curve $\varkappa\in (4,8)$, or a space filling Peano curve $\varkappa\geq 8$. At this step we deal with the chordal version of SLE. The complement to a SLE$_{\varkappa}$ hull in the upper half-plane $\mathbb H$ is a simply connected domain that is mapped conformally onto $\mathbb H$ by a holomorphic function $g(z,t)$ satisfying the equation
\begin{equation}\label{SLE}
\frac{dg}{dt}=\frac{2}{g(z,t)-\xi_t}, \quad g(z,0)=z,
\end{equation}
where $\xi_t=\sqrt{\varkappa}B_t$, and $B_t$ is a normalized Brownian motion with the diffusion constant $\varkappa$. The function $g(z,t)$ is expanded as $\displaystyle g(z,t)=z+\frac{2t}{z}+\dots$. The equation \eqref{SLE} is called the Schramm-L\"owner equation and was studied first in \cite{LSW1}--\cite{LSW3}, see also \cite{RohdeSchramm01} for basic properties of SLE.
Special values of $\varkappa$ correspond to interesting special cases of SLE, for example $\varkappa=2$ corresponds to the loop-erasing random walk and the uniform spanning tree, $\varkappa=4$ corresponds to the harmonic explorer and the Gaussian free field. Observe, that the equation \eqref{SLE} is not a stochastic differential equation (SDE). To rewrite it in a stochastic way (following \cite{BB}, \cite{FriedrichWerner}) let us set a function $k_t(z)=g(z,t)-\xi_t$, where $k_t(z)$ satisfies already the SDE
\[
dk_t(z)=\frac{2}{k_t(z)}dt-d\xi_t.
\]
For a function $F(z)$ defined in the upper half-plane one can derive the It\^o differential
\begin{equation}\label{Ito}
dF(k_t)=-d\xi_t L_{-1}F(k_t)+dt (\frac{\varkappa}{2}L_{-1}^2-2L_{-2})F(k_t), 
\end{equation}
with the operators $L_{-1}=-\frac{d}{d z}$ and $L_{-2}=-\frac{1}{z}\frac{d}{d z}$. These operators  are the first two Virasoro generators in the `negative' part of the Witt algebra spanned by the operators $-z^{n+1}\frac{d}{d z}$ acting on the appropriate representation space. For any state $|\psi\rangle$, the state $L_{-1}|\psi\rangle$ measures the diffusion of $|\psi\rangle$ under SLE, and
$(\frac{\varkappa}{2}L_{-1}^2-2L_{-2})|\psi\rangle$ measures the drift. The states of interest are drift-less, i.e., the second term in \eqref{Ito} vanishes. Such states are annihilated by $\frac{\varkappa}{2}L_{-1}^2-2L_{-2}$, which is true if we choose the state $|\psi\rangle$ as the highest weight vector  in the highest weight representation of the Virasoro algebra  with the central charge $c$ and the conformal weight $h$ given by
\[
c=\frac{(6-\varkappa)(3\kappa-8)}{2\varkappa},\quad h=\frac{6-\varkappa}{2\varkappa},
\]
and the operators $L_{-1}$ and $L_{-2}$ are  taken in the corresponding representation. It was obtained in  \cite{BB} and \cite{FriedrichWerner}, that $F(k_t)$ is a martingale if and only if
 $(\frac{\varkappa}{2}L_{-1}^2-2L_{-2})F(k_t)=0$.
We define a CFT with a boundary in $\mathbb H$ such that the boundary condition is changed by a boundary operator. The random curve in $\mathbb H$ defined by SLE is growing so that
it has states of one type to the left and of the other type to the right (the simplest way to view this is the lattice Ising model with the states defined as spin positions up or down). The mapping $g$ satisfying \eqref{SLE} `unzips' the boundary. The primary operator that induces the boundary change
with the conformal weight $h$ is drift-less, and therefore, its expectation value does not change in time under the boundary unzipping. Hence all correlators computing with this operator remain invariant. Analogous considerations one may provide for the `radial' version of SLE in the unit disk, slightly modifying the above statements.

Observe that in this formulation two Virasoro generators can generate the non-trivial `negative' part of the Witt algebra by the commutation relation
\[
[L_m,L_n]=(n-m)L_{n+m}.
\]

There are many forthcoming directions that can follow the described study of the L\"ownere-Kufarev equations. One of possible directions is to study
the sub-Riemannian geometry naturally related to the bracket generating structure of the Virasoro algebra and to the hypoellipticity of the drift operator for SLE. Another is to consider analogues of SLE
in the case of infinite degrees of freedom (stochastic version of the L\"oewner-Kufarev equation).  The alternate L\"owner-Kufarev equation, infinite dimensional controllable systems analogous to one considered here, are also new objects to study. We hope this survey will encourage a new wave of interest to this classical subject. 


\end{document}